\documentclass{emulateapj}

\usepackage{color}
\usepackage{natbib}
\usepackage{amssymb,amsmath}
\usepackage[dvipdfm,
bookmarks=true,                   
bookmarksnumbered=true,  
colorlinks=true,      
citecolor=blue,                     
linkcolor=blue,              
menucolor=blue,                
urlcolor=cyan,                       
breaklinks,
frenchlinks=False]{hyperref}

\newcommand\eex[1]{\mbox{$\times 10^{#1}$}}     % #1 x 10^#2
\newcommand\eez[1]{\mbox{$10^{#1}$}}            % just 10^#2

\newcommand\lsun{\hbox{$L_{\odot}$}}
\newcommand\msun{\hbox{$M_{\odot}$}}

\newcommand\kps{\mbox{${\rm km~s^{-1}}$}}
\newcommand\cm{\mbox{${\rm cm^{-2}}$}}
\newcommand\nh{\mbox{$N_{\rm HI}$}}

\newcommand{\bc}{\begin{center}}
\newcommand{\ec}{\end{center}}

\newcommand{\rc}{\mbox{$r_{c}$}}

\newcommand{\ergss}{erg~s$^{-1}$}
\newcommand{\ergscms}{erg~cm$^{-2}$~s$^{-1}$}

\newcommand{\rth}{\mbox{$r_{200}$}}
\newcommand{\rfh}{\mbox{$r_{500}$}}

\newcommand{\mth}{\mbox{$M_{200}$}}
\newcommand{\mfh}{\mbox{$M_{500}$}}

\newcommand{\chandra}{\textit{Chandra\/}}
\newcommand{\xmm}{\textit{XMM-Newton\/}}
\newcommand{\rosat}{\textit{ROSAT\/}}

\slugcomment{Accepted by the Astrophysical Journal, 20 December 2011}

\shorttitle{Finding Fossil Groups}
\shortauthors{Miller et al. 2011}

\begin{document}

\title{Finding Fossil Groups: Optical Identification and X-ray Confirmation}

\author{Eric D. Miller\altaffilmark{1},
Eli S. Rykoff\altaffilmark{2},
Renato A. Dupke\altaffilmark{3,4,5},
Claudia Mendes de Oliveira\altaffilmark{6},\\
Raimundo Lopes de Oliveira\altaffilmark{7,8},
Robert N. Proctor\altaffilmark{6},
Gordon P. Garmire\altaffilmark{9},\\
Benjamin P. Koester\altaffilmark{10},
Timothy A. McKay\altaffilmark{11}}

\altaffiltext{1}{Kavli Institute for Astrophysics and Space Research,
Massachusetts Institute of Technology, 77 Massachusetts Ave., Cambridge, MA
02139; milleric@mit.edu}
\altaffiltext{2}{E.O.~Lawrence Berkeley National Lab, 1 Cyclotron Rd.,
Berkeley, CA 94720}
\altaffiltext{3}{Department of Astronomy, University of Michigan, 500
Church St., Ann Arbor, MI 48109}
%\altaffiltext{4}{National Observatory, Rio de Janeiro, Brazil}
\altaffiltext{4}{Observat\'orio Nacional, Rua Gal.~Jos\'e Cristino 77, S\~ao Crist\'ov\~ao, CEP20921-400 Rio de Janeiro RJ, Brazil}
\altaffiltext{5}{Eureka Scientific Inc., 2452 Delmer St.~Suite 100, Oakland, CA 94602}
\altaffiltext{6}{Departamento de Astronomia, Instituto de Astronomia, Geof\'isica e Ci\^encias Atmosf\'ericas da Universidade de S\~ao Paulo, Rua do Mat\~ao 1226, Cidade Universit\'aria, 05508-090 S\~ao Paulo, Brazil }
\altaffiltext{7}{Universidade de S\~ao Paulo, Instituto de F\'isica de S\~ao Carlos, Caixa Postal 369, 13560-970 S\~ao Carlos, SP, Brazil}
\altaffiltext{8}{Universidade Federal de Sergipe, Departamento de F\'isica, Av. Marechal Rondon s/n, 49100-000 S\~ao Crist\'ov\~ao, SE, Brazil}
\altaffiltext{9}{Department of Astronomy and Astrophysics, Pennsylvania State University, 525 Davey Lab, University Park, PA 16802, USA}
\altaffiltext{10}{Department of Astronomy and Astrophysics, The University of Chicago, Chicago, IL 60637}
\altaffiltext{11}{Department of Physics, University of Michigan, 450 Church St., Ann Arbor, MI 48109}

\begin{abstract}
We report the discovery of 12 new fossil groups of galaxies, systems
dominated by a single giant elliptical galaxy and cluster-scale
gravitational potential, but lacking the population of bright galaxies
typically seen in galaxy clusters.  These fossil groups (FGs), selected
from the maxBCG optical cluster catalog, were detected in snapshot
observations with the \chandra\ X-ray Observatory.  We detail the highly
successful selection method, with an 80\% success rate in identifying
12 FGs from our target sample of 15 candidates.  For 11 of the systems,
we determine the X-ray luminosity, temperature, and hydrostatic mass, which
do not deviate significantly from expectations for normal systems, spanning
a range typical of rich groups and poor clusters of galaxies.  A small
number of detected FGs are morphologically irregular, possibly due to past
mergers, interaction of the intra-group medium (IGM) with a central AGN, or
superposition of multiple massive halos.  Two-thirds of the X-ray-detected
FGs exhibit X-ray emission associated with the central BCG, although we are
unable to distinguish between AGN and extended thermal galaxy emission
using the current data.  
This sample, a large increase in the number of known FGs, will be
invaluable for future planned observations to determine FG temperature,
gas density, metal abundance, and mass distributions, and to compare
to normal (non-fossil) systems.  Finally, the presence of a population of
galaxy-poor systems may bias mass function determinations that measure
richness from galaxy counts.  When used to constrain power spectrum
normalization and $\Omega_m$, these biased mass functions may in turn bias
these results.  
\end{abstract}


\section{INTRODUCTION}
\label{sect:intro}

Fossil groups (FGs) are systems dominated by a single, giant elliptical
galaxy, yet their X-ray emission indicates a deeper cluster-scale
gravitational potential.  
They are generally defined as systems with
a $\Delta R$ = 2 magnitude difference between the first and second rank
galaxies within $0.5\,$\rth\footnote{\texorpdfstring{$r_{200}$ is the radius within which the mean cluster mass density is 200 times the critical density.}{r200 is the radius within which the mean cluster mass density is 200 times the critical density.}},
and they have an extended thermal X-ray halo with 
$L_{X,{bol}} > \eez{42}\,h_{50}^{-2}$ \ergss\ \citep{Jonesetal2003}. 
FGs are thought to be old,
isolated galaxy groups and clusters in which the large galaxies have
coalesced through dynamical friction
\citep{Ponmanetal1994,MulchaeyZabludoff1999,Jonesetal2003}.
This coalesced cluster scenario is further supported by high X-ray
temperature measurements (up to $\sim$ 4 keV) and 
by the galaxy velocity dispersions  \citep[e.g.,][]{MendesdeOliveiraetal2006,MendesdeOliveiraetal2009,Cyprianoetal2006,Proctoretal2011}. 
The high NFW \citep{NFW1997} halo concentration parameters, lack of
spectral star formation indicators, and large $\Delta R$ magnitude
difference suggest these systems finished merging in the distant past,
perhaps before $z \sim 1$
\citep{Jonesetal2000,Wechsleretal2002,DOnghiaetal2005,Khosroshahietal2007}.

Recent studies of X-ray selected FGs paint a more complicated picture.  
The cooling time of FGs is significantly shorter than the Hubble time
\citep[e.g.,][]{Sunetal2004,Khosroshahietal2004,Khosroshahietal2006}, 
yet they typically lack cool cores, suggesting that these systems may be
younger or more active than previously thought
\citep{MendesdeOliveiraetal2009}.  Regular
(non-fossil) rich groups often possess cool cores
\citep[e.g.,][]{FinoguenovPonman1999}, even in the presence of AGN
activity.
In addition, there is evidence for enhanced SN II metal fraction in the
central regions of FGs, 
suggesting a scenario where SN II powered winds resulting from
merging late type galaxies erase the original central SN Ia Fe mass
fraction dominance 
\citep{Dupkeetal2010}.
This is consistent with the previously found disky isophotes of the
central dominant galaxies in FGs by \citet{KhosroshahiPonmanJones2006}
and also with the presence of shells in the stellar component in at least
one of these galaxies, indicative of multiple past mergers
\citep{EigenthalerZeilinger2009}.
Furthermore, some authors suggest that the FGs we see are the tail of the
cluster distribution, possessing few $L_\star$ galaxies at their current
epoch for any number of reasons: failure to form those galaxies, early
merging, or a quiescent state during a cycle of galaxy accretion
\citep{MulchaeyZabludoff1999,vonBendaBeckmannetal2008,laBarberaetal2009,Dariushetal2010,Smithetal2010,Cuietal2011}.
The truth about fossil groups is somewhat muddled by the phenomenological
rather than physical definition of the class.  Extracting a useful physical
definition is in turn complicated by the relatively small number of FGs
with deep X-ray observations.  Indeed, what we call ``fossil groups''
perhaps comprise a heterogeneous set of galaxy systems with different
formation and evolution histories.

To characterize the ages and structural properties of FGs, it is crucial to
have good data, especially X-ray observations.  Detailed study of the
intra-group medium (IGM)
metal abundance, temperature structure, and inferred mass distribution help
to constrain the halo formation epoch and the importance of recent star
formation or AGN activity.  The available X-ray data are typically
photon-poor due to the serendipitous nature of FG detections, and this has
limited their study.  To address this problem, we have
embarked on a project to identify a large sample of FGs for
future detailed follow-up studies.  In the work presented here, we have
constructed a sample of 15 fossil group candidates, using the maxBCG
cluster catalog \citep{Koesteretal2007b} to optically identify the
candidates, which are then targeted with X-ray snapshots using the
\chandra\ X-ray Observatory to confirm the existence of a bright X-ray
halo.  This initial sample of confirmed FGs (using the
phenomenological classification) will be invaluable for follow-up,
including deep X-ray observations to study the metallicity structure of
the gas and concentration of the mass distribution; and optical
spectroscopy to compare the velocity dispersion to the X-ray mass.

Throughout this paper we use a $\Lambda$CDM cosmology with
$\Omega_m = 0.27$, $\Omega_{\Lambda} = 0.73$, and $H_0 = 70$ km s$^{-1}$
Mpc$^{-1}$ (or $h = 0.7$).  Except where specified with $h$ notation, all
numerical values from the literature have been scaled to correspond to this
cosmology.  Uncertainties are 1$\sigma$ and upper/lower limits are
3$\sigma$, unless stated otherwise.

\section{OPTICAL SAMPLE SELECTION}
\label{sect:opt}

To select a large sample of FG targets, we used the maxBCG cluster catalog
\citep{Koesteretal2007b}.  This is a volume-limited catalog of over 17,000
optically selected red-sequence clusters in the redshift range 
$0.1 < z < 0.3$
with precise photometric redshifts ($\delta_z\sim0.01$) and optical
richness estimates from the Sloan Digital Sky Survey DR4
\citep[SDSS;][]{Yorketal2000,Adelman-McCarthyetal2006}.  The optical
richness employed, $N_{200}$, is the number of red-sequence member galaxies
brighter than $0.4\,L_*$ (in the $i$-band) found within a scale radius
\rth\ of the
brightest cluster galaxy \citep[BCG;][]{Hansenetal2005}. The mean
properties of this catalog have been studied in detail, and we have
obtained mean X-ray luminosities~\citep{Rykoffetal2008}, velocity
dispersions~\citep{Beckeretal2007}, and a mass calibration via the mean
weak-lensing shear profile around maxBCG clusters
\citep{Johnstonetal2007,Sheldonetal2009}.  Simulations have shown that the
catalog purity and completeness are very high
\citep[$>$90\textrm{\%};][]{Koesteretal2007a,Rozoetal2007a}. 

We selected for several optical characteristics that are expected of FGs
from the empirical definition of the class.
In particular, at a given optical richness, these systems
should have a larger magnitude difference between the BCG and the next
brightest galaxy and should have highly luminous BCGs typical of massive
clusters.  In our initial selection we restricted ourselves to systems in
the richness range $9 \le N_{200} \le 25$, corresponding to a mass range of
$3\eex{13} \lesssim M_{200} \lesssim 1\eex{14}\,h^{-1}\,M_\sun$
\citep{Johnstonetal2007}.  Our aim was to select systems rich enough to
have sufficient X-ray luminosity for detection and analysis, and at the
same time remain in the FG range where the BCG can truly dominate the
system.  We note that the publicly released maxBCG catalog was restricted
to $N_{200} \ge 10$; we used a slightly extended catalog described by
\citet{Rykoffetal2008}, including an additional 3532 clusters with $N_{200} =
9$ and allowing us to use the same richness bins studied in the analysis of
maxBCG galaxy dynamics \citep{Beckeretal2007} and weak lensing
\citep{Johnstonetal2007,Sheldonetal2009}.
To maximize the flux in the X-ray, we restricted the sample to those
systems with a confirmed spectroscopic BCG redshift in the range
$0.09 \leq z \leq 0.15$; the lower redshift cut is imposed by
the maxBCG photometric redshift lower limit of $z = 0.10 \pm 0.01$.
At the time of selection from SDSS DR4, 42\% of all maxBCG
clusters in this redshift range had a spectroscopically determined BCG
redshift.

\begin{figure}
\centering
\includegraphics[width=.9\linewidth]{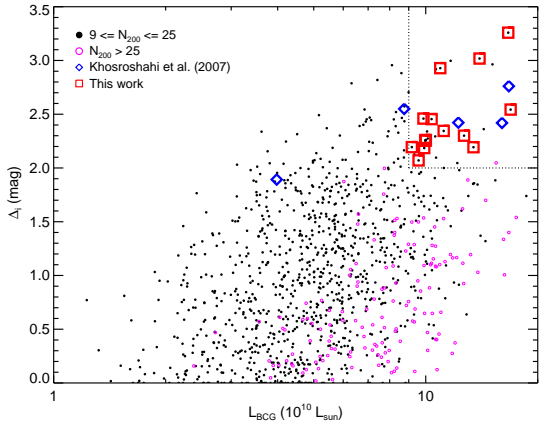}
\caption{Difference in $i$-band magnitude ($\Delta_i$) of the BCG and next
brightest red-sequence cluster member within $0.5\,\rth$ as a function of
BCG $i$-band luminosity ($L_{BCG}$) for all maxBCG systems with measured
spectroscopic BCG redshifts $0.09 \leq z \leq 0.15$.  Our FG candidates (red
squares) were chosen from the subset of systems with $9 \le N_{200} \le 25$
(black dots) within the region delineated by the dotted lines; for
comparison we also show all systems with $N_{200} > 25$ (open magenta
circles).  Known FGs from \citet{Khosroshahietal2007} that overlap the SDSS
footprint and are detected by maxBCG (blue diamonds) are shown for
illustration although they do not meet our selection criteria.  Note that
three of our candidate FGs have very similar values for $\Delta_i$
(2.25--2.26) and $L_{BCG}$ ($10\eex{10}$ \lsun), so the red squares
overlap.}
\label{fig:selection}
\end{figure}

To quantify the galaxy magnitude gap representative of FGs, we used
the difference in $i$-band magnitude of the BCG and the next
brightest red-sequence cluster member within $0.5\,\rth$, denoted by
$\Delta_i$.  The $\rth$ values were estimated from the mass-scaling
relation of \citet{Johnstonetal2007}.
Bright non-red-sequence galaxies projected within $0.5\,\rth$ were not
considered because the majority of these are foreground galaxies
unassociated with the cluster.  Thus, we have opted for a more complete
sample of systems with this magnitude gap at the risk of a small amount of
impurity.  We refined our selection to all the systems with $\Delta_i >
2.0$ and BCG $i$-band luminosity $L_{BCG} > 9\times10^{10}\,h^{-2}\,L_\sun$
($L_{BCG} > 1.8\times10^{11}\,L_\sun$ in our adopted cosmological
framework).  There were 26 maxBCG systems that passed this selection cut.
We illustrate the maxBCG selection in Figure~\ref{fig:selection}, plotting
$\Delta_i$ as a function of $L_{BCG}$; these two values are correlated, as
the most luminous BCGs also tend to have a large $\Delta_i$.  However,
richer maxBCG clusters ($N_{200} > 25$) in the same redshift range (magenta
points) do not extend this trend to the largest $\Delta_i$, despite having
relatively larger $L_{BCG}$.  We discuss selection effects of our sample
more fully in a companion paper by \citet{Proctoretal2011}.  The cuts based
on these parameters are consistent with the majority of known FG
systems~\citep[e.g.][]{Khosroshahietal2007} that overlap the SDSS
footprint, marked with blue diamonds in Figure~\ref{fig:selection}.  These
five systems are present in the maxBCG catalog, however they each fail one
or more of our selection criteria, with four falling out of our redshift
range and the fifth having $N_{200} = 28$.

Deeper X-ray follow-up is a major goal of this sample assembly, and the
most efficient current instrument for this purpose is the \xmm\ X-ray
Observatory.  The $\sim$ 15 arcsec spatial resolution of \xmm\ corresponds
to 30 kpc at $z = 0.1$.  This is a sizeable fraction of the typical group
core radius, so to maximize the utility of these targets for deeper X-ray
spectroscopic analysis we sought to reduce the possibility of bright
central AGN emission that might contaminate the diffuse IGM emission.  We
rejected all BCGs that have evidence of Seyfert or LINER-like line emission
with $\log([\mathrm{NII}]/\mathrm{H}\alpha)>-0.2$ \citep[e.g.][]{khtbc03}.
This eliminated 3 of the 26 systems.  Additionally, we rejected all BCGs
that match radio sources in the FIRST catalog \citep{wbhg98} within 3
arcsec of the BCG position (6 kpc at $z = 0.1$), eliminating another 6
candidates. 
Finally, we rejected 2 systems that have bright stars with
$m_R<7$ in the field-of-view that exceed the bright star limit for \xmm.
In total, we identified 15 candidate FGs, marked with red squares in
Figure~\ref{fig:selection} and summarized in Table~\ref{tab:obs}.

\begin{figure}
\centering
\includegraphics[width=\linewidth]{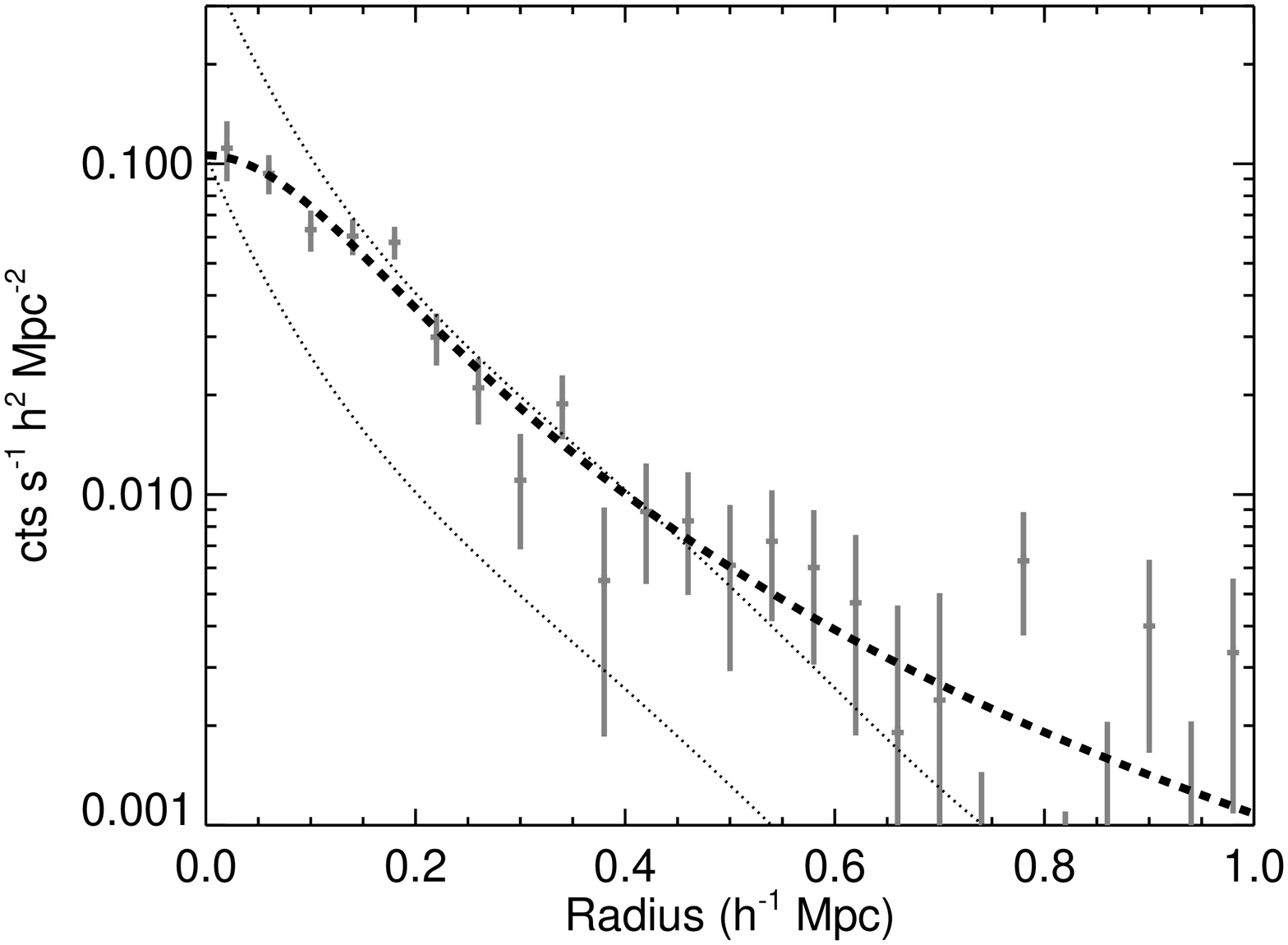}
\caption{Radial stacked RASS profile of 170 maxBCG clusters with a similar
selection function to those targeted in this paper.  The thick dashed line
shows the best-fit $\beta$-model surface brightness profile, which yields a
good fit with $\beta = 0.6\pm0.1$.  The dotted lines show profiles for a
RASS point source with two normalizations; the lower one is scaled to the
central surface brightness, while the higher one is scaled to the surface
brightness at $0.4\,h^{-1}\,\mathrm{Mpc}$.  In both cases it is clear that
the stacked X-ray emission of this sample is significantly extended when
compared to a RASS point source.}
\label{fig:stackprofile}
\end{figure}

The elimination of candidates based on AGN activity, especially in the
radio, could produce a selection bias.  The motivation for this selection
was based on \citet{Allenetal2006} and \citet{Balmaverdeetal2008}, who find
a clear correlation between Bondi accretion rate and central engine jet
power in samples of low-power radio galaxies.  The presence of radio
emission from a jet or expanding bubbles is a signal of strong accretion,
with rate $\dot M_{Bondi} \propto \rho$.  As the X-ray emission measure of
this plasma is $EM \propto \rho^{2}$, the flux within the central few kpc
can be quite bright and contaminate the \xmm\ surface brightness profile,
and through an abundance of caution, we expunged these radio-bright
candidates.  The direct correlation between AGN radio and X-ray luminosity
in BCGs is not well constrained, although \citet{Hickoxetal2009} find a
small overlap ($\sim$ 10\%) between radio- and X-ray-bright AGN in general,
with the former tending to live in luminous red sequence galaxies and the
latter tending to ``green valley'' galaxies.
The possible effects of this selection on our sample are discussed in
Section~\ref{sect:success}.

The final criterion in the FG definition is a cluster-scale X-ray halo,
which also serves as confirmation of a collapsed system rather than a
projection of unrelated galaxies. 
While six of the FG candidates are bright enough to be detected at the
$>2\sigma$ level in the \rosat\ All Sky Survey (RASS), a majority of
these optically selected targets do not have sufficient flux to be detected
in RASS (see Table~\ref{tab:obs}).  We therefore estimated the mean $L_X$
and radial profile using the stacking procedure described by
\citet{Rykoffetal2008},
selecting 170 maxBCG systems chosen with criteria used for the 15 FG
targets ($9 \le N_{200} \le 25$; $\Delta_i > 2.0$; and
$L_{BCG} > 9\times10^{10}\,L_\sun$).  In order to select enough systems for
the stacking analysis we expanded the redshift range to 
$0.09 \leq z \leq 0.20$, and we did not require BCG spectroscopic redshifts
nor did we filter for active galaxies or bright stars.  
The stacked RASS profile is shown in
Figure~\ref{fig:stackprofile} along with a $\beta$ model fit with 
$\beta = 0.6 \pm 0.1$.  To show that the stacked FG profile is
significantly extended, we also stacked a representative sample of RASS
point sources treated as if these sources were at the redshifts of the
maxBCG targets~\citep[see Section~3.3.1 in][]{Rykoffetal2008}.  The stacked
point source profile is shown with dotted lines with two scalings in
Figure~\ref{fig:stackprofile}.  In one scaling the central flux is matched
to the stacked FG profile; in the second, the flux at $0.4\,\mathrm{Mpc}$
is matched to the FG profile.  In both cases it is clear that the stacked
FG emission is significantly extended when compared to RASS point sources.  

\begin{deluxetable*}{ccccccccccc}
\tabletypesize{\footnotesize}
\tablewidth{0pt}
\tablecaption{Fossil Group Sample and \chandra\ Observations
     \label{tab:obs}}
\tablehead{
\colhead{Target BCG} &
\colhead{short name\tablenotemark{a}} &
\colhead{RA\tablenotemark{b}} &
\colhead{Dec\tablenotemark{b}} &
\colhead{$z$\tablenotemark{c}} &
\colhead{$N_{200}$\tablenotemark{d}} &
\colhead{$\Delta_i$\tablenotemark{e}} &
\colhead{$L_{BCG}$\tablenotemark{f}} &
\colhead{RASS\tablenotemark{g}} &
\colhead{obs.~date (OBSID)} &
\colhead{$t_{exp}$\tablenotemark{h}} \\
\colhead{} &
\colhead{} &
\colhead{} &
\colhead{} &
\colhead{} &
\colhead{} &
\colhead{(mag)} &
\colhead{($\eez{10}$ \lsun)} &
\colhead{} &
\colhead{} &
\colhead{(ksec)} 
}
\startdata
\object[SDSS J013325.87-102618.6]{SDSS~J013325.87$-$102618.6} & J0133$-$1026 & \phn23.3578  &    $-$10.4385  & 0.113  &    12  & 2.45  &    10.4  & \nodata  & 2009-05-29 (10753) &    10.0 \\
\object[SDSS J081526.59+395935.5]{SDSS~J081526.59$+$395935.5} & J0815$+$3959 &    123.8608  &    $+$39.9932  & 0.129  &    12  & 3.26  &    16.7  & $\surd$  & 2008-12-12 (10758) & \phn5.1 \\
\object[SDSS J082122.54+405123.7]{SDSS~J082122.54$+$405123.7} & J0821$+$4051 &    125.3439  &    $+$40.8566  & 0.125  &    10  & 2.19  & \phn9.9  & \nodata  & 2009-01-04 (10474) &    10.0 \\
\object[SDSS J085640.72+055347.3]{SDSS~J085640.72$+$055347.3} & J0856$+$0553 &    134.1697  & \phn$+$5.8965  & 0.094  &    16  & 2.26  &    10.0  & \nodata  & 2009-01-09 (10750) & \phn5.5 \\
\object[SDSS J090638.27+030139.1]{SDSS~J090638.27$+$030139.1} & J0906$+$0301 &    136.6595  & \phn$+$3.0276  & 0.136  & \phn9  & 2.93  &    10.9  & \nodata  & 2009-01-14 (10475) &    10.0 \\
\object[SDSS J100742.53+380046.6]{SDSS~J100742.53$+$380046.6} & J1007$+$3800 &    151.9272  &    $+$38.0130  & 0.112  &    24  & 2.54  &    16.9  & $\surd$  & 2009-02-09 (10755) & \phn4.7 \\
\object[SDSS J101745.57+015645.8]{SDSS~J101745.57$+$015645.8} & J1017$+$0156 &    154.4399  & \phn$+$1.9461  & 0.118  &    12  & 2.34  &    11.2  & \nodata  & 2009-03-23 (10754) & \phn9.9 \\
\object[SDSS J103930.43+394718.9]{SDSS~J103930.43$+$394718.9} & J1039$+$3947 &    159.8768  &    $+$39.7886  & 0.093  &    14  & 2.46  & \phn9.9  & \nodata  & 2009-01-14 (10749) & \phn5.1 \\
\object[SDSS J104548.50+042032.5]{SDSS~J104548.50$+$042032.5} & J1045$+$0420 &    161.4521  & \phn$+$4.3424  & 0.154  &    13  & 2.07  & \phn9.6  & \nodata  & 2009-02-01 (10476) & \phn9.9 \\
\object[SDSS J113305.51+592013.7]{SDSS~J113305.51$+$592013.7} & J1133$+$5920 &    173.2730  &    $+$59.3372  & 0.133  &    13  & 2.26  &    10.0  & $\surd$  & 2009-07-08 (10472) & \phn5.7 \\
\object[SDSS J113623.71+071337.5]{SDSS~J113623.71$+$071337.5} & J1136$+$0713 &    174.0988  & \phn$+$7.2271  & 0.103  &    17  & 2.25  &    10.0  & $\surd$  & 2009-02-09 (10756) & \phn5.0 \\
\object[SDSS J115305.32+675351.5]{SDSS~J115305.32$+$675351.5} & J1153$+$6753 &    178.2722  &    $+$67.8977  & 0.117  &    17  & 2.19  &    13.4  & $\surd$  & 2009-06-21 (10473) & \phn5.0 \\
\object[SDSS J133626.96+545353.8]{SDSS~J133626.96$+$545353.8} & J1336$+$5453 &    204.1124  &    $+$54.8983  & 0.107  &    10  & 3.02  &    13.9  & \nodata  & 2009-09-25 (10752) & \phn7.1 \\
\object[SDSS J141004.19+414520.8]{SDSS~J141004.19$+$414520.8} & J1410$+$4145 &    212.5175  &    $+$41.7558  & 0.094  &    21  & 2.30  &    12.7  & $\surd$  & 2009-07-07 (10757) & \phn5.1 \\
\object[SDSS J141115.89+573609.0]{SDSS~J141115.89$+$573609.0} & J1411$+$5736 &    212.8162  &    $+$57.6025  & 0.106  &    16  & 2.19  & \phn9.2  & \nodata  & 2009-07-23 (10751) & \phn6.9 \\
\enddata
\tablenotetext{a}{The short names for each target are used throughout this work.}
\tablenotetext{b}{RA, Dec are the J2000 coordinates of the BCG, in degrees.}
\tablenotetext{c}{The redshift is the BCG spectroscopic value from SDSS DR4 \citep{Adelman-McCarthyetal2006}.}
\tablenotetext{d}{$N_{200}$, a richness estimate, is the number of red-sequence cluster galaxies brighter than $0.4\,L_*$ (in the $i$-band) found within $r_{200}$ of the BCG.}
\tablenotetext{e}{$\Delta_i$ is the difference in $i$-band magnitude of the BCG and the next brightest red-sequence cluster member within $0.5\,\rth$.}
\tablenotetext{f}{$L_{BCG}$ is the $i$-band luminosity of the BCG.}
\tablenotetext{g}{RASS indicates a 2$\sigma$ or better detection in the ROSAT All Sky Survey.}
\tablenotetext{h}{Effective exposure time of cleaned event data.}
\end{deluxetable*}

The derived $L_X$ estimates were used to plan our follow-up X-ray
observations with \chandra, described in the next Section.
\citet{Rykoffetal2008} show that the mean $L_X$ of maxBCG systems scales as
a power-law with $N_{200}$ (over two orders of magnitude in $L_X$) and also
scales with $L_{BCG}$ for the poorer clusters and groups.  Thus, the FGs
are expected to be more X-ray luminous than typical maxBCG systems at a
similar richness.

\section{X-RAY OBSERVATIONS AND DATA ANALYSIS}
\label{sect:obsandanalysis}

\subsection{X-Ray Data and Reduction}
\label{sect:data}

The X-ray observations were performed with the \chandra\ X-ray Observatory
between December 2008 and September 2009 in the form of 5--10 ksec
snapshots (see Table~\ref{tab:obs}).  Data were obtained using the ACIS-S3
chip, with the candidate fossil group BCG centered in the field of view.
Standard processing was performed on the raw event files, including the
background reduction tools applicable to the VFAINT observing
mode\footnote{See
\url{http://cxc.harvard.edu/cal/Acis/Cal\_prods/bkgrnd/current/background.html}.}.
The resulting 0.3--7 keV light curves were filtered to remove additional
times of high background and applied to produce cleaned event files.
Point sources were identified from the 0.3--7 keV events using the CIAO
tool {\tt wavdetect} and masked out for the analysis of the extended emission.
This included any emission clearly identified with the optical extent 
of the BCG, whether point-like or slightly extended.  A separate analysis
of these features is presented in Section~\ref{sect:ptsrc}.

\subsection{X-Ray Spectral Analysis}
\label{sect:spectral}

Spectral extraction regions were identified from images in the 0.5--2 keV
band, where the group emission should dominate, and chosen to encompass the
bulk of the extended X-ray emission, with radii in the range
$1.2\arcmin$--$2.2\arcmin$ (129--356 kpc) centered on the peak of the
X-ray emission.  For two targets with irregular X-ray morphology
(J0133$-$1026 and J1045$+$0420), the region was centered on the apparent
centroid of the emission.  Targets without obvious group emission
(J0821$+$4051, J0906$+$0301, and J1336$+$5453) were assigned an extraction
region of 250 kpc ($1.7\arcmin$--$2.1\arcmin$) centered on the optical BCG
location to estimate an upper limit on the flux.  A background region was
defined for each observation from the remaining area of the
$8.4\arcmin\times8.4\arcmin$ ACIS-S3 chip, excluding the inner $\sim
2.5\arcmin$ radius, the identified point sources, and the outer edge of the
field of view.  The outer extent of each background region was typically
$7.5\arcmin \times 5.5\arcmin$ in size, oriented along the direction of the
ACIS-S CCD array.  This choice was made to reduce the effects of
non-uniform molecular contamination on the CCD, which is thought to be
thicker near the edges of the ACIS-S array and which substantially reduces
the soft X-ray transmittance \citep{Vikhlinin2004}.

\begin{figure*}[t]
\centering
\includegraphics[height=\linewidth]{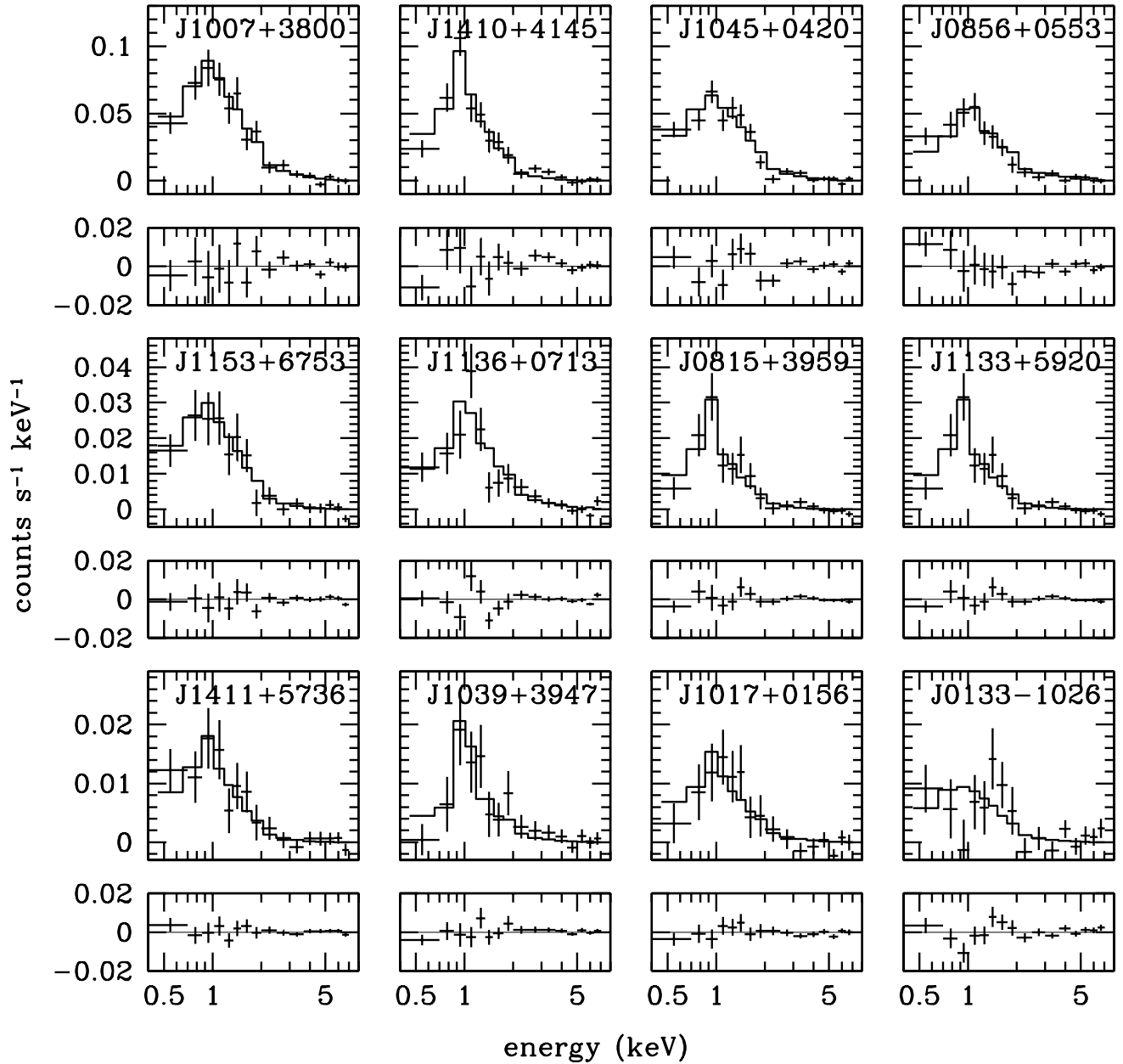}
\caption{Background-subtracted \chandra/ACIS-S3 spectra of the 12 detected
FGs with the best-fit models.  The displayed spectra have been binned in
energy for clarity, although the spectral analysis was carried out on the
full resolution (14.6 eV/channel) ACIS spectra.  Shown below each spectrum
are the fit residuals.  The FGs are shown roughly in order of brightest to
faintest for clarity of the ordinate scaling.}
\label{fig:spec}
\end{figure*}

The spectral analysis was performed in XSPEC v12.6.0 utilizing the
C-statistic, a modified \citet{Cash1979} likelihood function that allows
for inclusion of a background spectrum and a goodness-of-fit estimator
similar to $\chi^2$ in the limit of many counts\footnote{The XSPEC
implementation of the C-statistic is described in detail at
\url{http://heasarc.gsfc.nasa.gov/docs/xanadu/xspec/manual/XSappendixCash.html}.}.
The X-ray spectrum for each group (see Figure~\ref{fig:spec}) was fit with
an absorbed APEC model \citep{Smithetal2001} in the 0.4--7 keV band, with
the redshift fixed at the BCG spectroscopic value from SDSS.  The
intervening Galactic \nh\ column was fixed at the average value reported by
the Leiden/Argentine/Bonn (LAB) merged survey \citep{Kalberlaetal2005}; in
all cases it was less than $5\times 10^{20}$ \cm.  The temperature and
normalization were allowed to vary.  For several groups, the metal
abundance (using \citet{AndersGrevesse1989} photospheric solar abundances)
was unconstrained during the initial fit and was frozen at 0.3 solar, the
weighted mean value from the well-constrained fits and similar to that
measured in $\sim$ 2 keV systems
\citep{OsmondPonman2004,RasmussenPonman2007}.  The spectral fitting results
are shown in Table~\ref{tab:spectral}.

\begin{figure*}[t]
\centering
\begin{minipage}[t]{.24\linewidth}
\centering
\includegraphics[width=\linewidth]{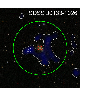}
\end{minipage}
\begin{minipage}[t]{.24\linewidth}
\centering
\includegraphics[width=\linewidth]{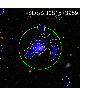}
\end{minipage}
\begin{minipage}[t]{.24\linewidth}
\centering
\includegraphics[width=\linewidth]{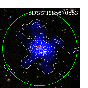}
\end{minipage}
\begin{minipage}[t]{.24\linewidth}
\centering
\includegraphics[width=\linewidth]{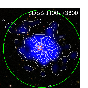}
\end{minipage}
\\
\begin{minipage}[t]{.24\linewidth}
\centering
\includegraphics[width=\linewidth]{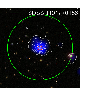}
\end{minipage}
\begin{minipage}[t]{.24\linewidth}
\centering
\includegraphics[width=\linewidth]{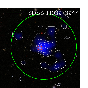}
\end{minipage}
\begin{minipage}[t]{.24\linewidth}
\centering
\includegraphics[width=\linewidth]{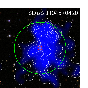}
\end{minipage}
\begin{minipage}[t]{.24\linewidth}
\centering
\includegraphics[width=\linewidth]{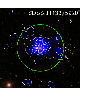}
\\
\end{minipage}
\begin{minipage}[t]{.24\linewidth}
\centering
\includegraphics[width=\linewidth]{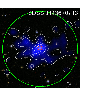}
\end{minipage}
\begin{minipage}[t]{.24\linewidth}
\centering
\includegraphics[width=\linewidth]{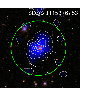}
\end{minipage}
\begin{minipage}[t]{.24\linewidth}
\centering
\includegraphics[width=\linewidth]{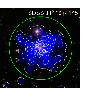}
\end{minipage}
\begin{minipage}[t]{.24\linewidth}
\centering
\includegraphics[width=\linewidth]{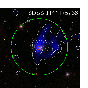}
\end{minipage}
\caption{SDSS multi-color images of the 12 \chandra-detected FGs, with the
diffuse 0.5--2 keV X-ray emission overlaid in contours.  The \chandra\
contours have point sources removed and are exposure-corrected and smoothed
with a 30\arcsec\ FWHM Gaussian.  The displayed blue intensity and contours
start at 2$\sigma$ (about 5\eex{-9} ph s$^{-1}$ cm$^{-2}$) above the
background surface brightness, and increase in intervals of 2$\sigma$.
Red circles note the optically identified BCG, and the green circle
indicates 0.5\,\rfh, centered on the X-ray peak (for well-defined
$\beta$-model fits) or BCG position.  The green bar in the upper left of
each panel shows 100 kpc at the FG redshift.  Images are 6\arcmin\ on a
side, ranging from 0.62 to 0.97 Mpc for the nearest and farthest FG,
respectively.}
\label{fig:smimages}
\end{figure*}

We consider a detection to be a 3$\sigma$ or greater excess of counts in
the 0.5--2 keV band compared to the expected background within the
extraction region.  Twelve of the 15 targets were detected using this
definition; all of the detections are easily visible to the eye in
point-source-excluded images (see Figure~\ref{fig:smimages}).  Eleven of
these detected systems have well-constrained temperatures in the range 1--3
keV; while J0133$-$1026 is detected at 4.9$\sigma$, the 104 source counts
are insufficient to constrain the spectral model.  Errors on the spectral
parameters were determined by sampling parameter space for each parameter,
marginalizing over the other free parameters.  For the systems with fixed
abundance, we estimated the errors in other parameters by stepping (with
XSPEC {\tt steppar}) the abundance over the expected range of 0.1--1 solar.
Absorbed fluxes and unabsorbed ``soft'' (0.5--2.0 keV rest frame)
luminosities were determined from the best-fit spectral models.  A
``bolometric'' (0.008--100 keV rest frame) luminosity with errors was
determined for each group by extrapolating the unabsorbed spectral model.
We assumed $kT = 2$ keV and abundance of 0.3 solar for the three undetected
targets, and used this model along with the background counts estimate to
determine upper limits on the diffuse X-ray flux and luminosity.  An
identical model was assumed for J0133$-$1026 to calculate its flux and
luminosity.  The source and total counts are shown in
Table~\ref{tab:spectral}, along with the flux and luminosity estimates.

\subsection{X-Ray Spatial Analysis}
\label{sect:spatial}

To compare to existing results and expectations for self-similar scaling,
cluster X-ray luminosities and masses are typically scaled to a common
radius in terms of the average interior overdensity $\delta_r =
\rho_r/\rho_{crit}$, where $\rho_r$ is the mean cluster mass density within
radius $r$, and $\rho_{crit}$ is the critical density at $z$
\citep[e.g.,][]{Maughanetal2006}.  The extraction regions we have applied
are considerably smaller than the typical radius for $\delta_r = 500$ for
group potentials ($\rfh \sim 500$ kpc, compared to $r_{extract} \sim 250$
kpc; see Table~\ref{tab:spatial}).  Therefore we must estimate the
luminosity correction factor by extrapolating the X-ray surface brightness
profile, similar to the approach in previous studies
\citep[e.g.,][]{Jeltemaetal2006}.  In the following analysis, we use the
working assumption that the intra-group medium is spherical, non-rotating,
isothermal, and in hydrostatic equilibrium to \rfh\ within the group
gravitational potential.

Counts images of the ACIS-S3 field of view were constructed from the
cleaned event lists, binning to $4\times4$ pixels ($2\times2$ arcsec) and
restricting the energy band to 0.5--2 keV, where the group emission
dominates the background.  For each detected group, two-dimensional spatial
fitting was performed with the Sherpa package available in CIAO, using a
circular $\beta$ model surface brightness profile for the FG emission and a
constant baseline to account for the combined cosmic and instrumental
background.  The $\beta$ model was multiplied by an exposure map during the
fit, while the background was multiplied by a mask containing the bad
pixels and columns of the CCD, dithered according to the aspect solution of
the observation.  Fits were performed in two dimensions to a region $\sim
7.5\arcmin \times 5.5 \arcmin$ oriented along the direction of the ACIS-S
array, encompassing the region defined as the background for spectral
fitting (see Section~\ref{sect:spectral}), chosen in such a way as to to
reduce the effects of non-uniform molecular contamination on the CCD.
Point source regions were excluded in the fitting, which was done using the
\citet{Cash1979} statistic, allowing $\beta$, core radius \rc, the emission
center, and the FG and background amplitude to vary.  The fit results are
summarized in Table~\ref{tab:spatial}.

\begin{deluxetable*}{ccccccccc}
\tabletypesize{\footnotesize}
\tablewidth{0pt}
\tablecaption{Spectral Fitting Results
\label{tab:spectral}}
\tablehead{
\colhead{FG} &
\colhead{\nh} &
\colhead{$kT$\tablenotemark{a}} &
\colhead{abund\tablenotemark{b}} &
\colhead{flux\tablenotemark{c}} &
\colhead{$L_{soft}$\tablenotemark{d}} &
\colhead{$L_{bol}$\tablenotemark{e}} &
\colhead{src/tot\tablenotemark{f}} &
\colhead{sig.\tablenotemark{g}} \\
\colhead{} &
\colhead{(\eez{20} \cm)} &
\colhead{(keV)} &
\colhead{(solar)} &
\colhead{(\eez{-14} cgs)} &
\colhead{(\eez{42} cgs)} &
\colhead{(\eez{42} cgs)} &
\colhead{counts} &
\colhead{($\sigma$)}
}
\startdata
J0133$-$1026 &  3.2 & 2.00                       &  0.30                   &  \phn3.4$^{+1.0}_{-1.0}$  &  \phn1.2$^{+0.4}_{-0.4}$   &  \phn2.9$^{+0.9}_{-0.8}$   &  104/455\phn    & \phn4.9   \\
J0815+3959   &  4.5 & 1.26$^{+0.18}_{-0.19}$     &  0.30                   &  \phn6.8$^{+1.3}_{-1.4}$  &  \phn3.5$^{+0.7}_{-0.7}$   &  \phn7.3$^{+1.4}_{-1.6}$   &  \phn87/308\phn & \phn5.0   \\
J0821+4051   &  4.6 & 2.00                       &  0.30                   &  $<$ 7.6                  &  $<$ 3.5                   &  $<$ 8.4                   &  \phn$\cdots$/742\phn & $\cdots$   \\
J0856+0553   &  3.6 & 2.73$^{+1.10}_{-0.57}$     &  0.30                   &  20.0$^{+2.1}_{-2.1}$     &  \phn4.8$^{+0.5}_{-0.5}$   &  12.6$^{+1.3}_{-1.3}$      &  283/498\phn    & 12.7  \\
J0906+0301   &  2.9 & 2.00                       &  0.30                   &  $<$ 3.8                  &  $<$ 2.0                   &  $<$ 4.8                   &  \phn$\cdots$/226\phn & $\cdots$   \\
J1007+3800   &  1.4 & 2.60$^{+0.63}_{-0.53}$     &  0.24$^{+0.32}_{-0.20}$ &  33.3$^{+3.6}_{-6.3}$     &  11.0$^{+1.2}_{-2.1}$      &  28.2$^{+3.0}_{-5.3}$      &  396/598\phn    & 16.2  \\
J1017+0156   &  3.9 & 2.13$^{+1.07}_{-0.55}$     &  0.30                   &  \phn4.7$^{+0.9}_{-0.8}$  &  \phn1.9$^{+0.4}_{-0.3}$   &  \phn4.5$^{+0.9}_{-0.8}$   &  118/400\phn    & \phn5.9   \\
J1039+3947   &  1.6 & 1.68$^{+0.86}_{-0.33}$     &  0.30                   &  \phn4.8$^{+1.2}_{-1.0}$  &  \phn1.1$^{+0.3}_{-0.2}$   &  \phn2.5$^{+0.6}_{-0.5}$   &  \phn63/123\phn & \phn5.7   \\
J1045+0420   &  3.5 & 2.47$^{+0.64}_{-0.47}$     &  0.11$^{+0.24}_{-0.11}$ &  24.5$^{+3.1}_{-3.1}$     &  17.2$^{+2.2}_{-2.2}$      &  43.7$^{+5.5}_{-5.6}$      &  610/1023       & 19.1  \\
J1133+5920   &  0.9 & 1.57$^{+0.47}_{-0.27}$     &  0.41$^{+0.40}_{-0.23}$ &  \phn7.1$^{+0.7}_{-1.6}$  &  \phn3.4$^{+0.3}_{-0.7}$   &  \phn7.5$^{+0.8}_{-1.6}$   &  110/184\phn    & \phn8.1   \\
J1136+0713   &  3.3 & 2.64$^{+1.43}_{-0.60}$     &  0.30                   &  10.1$^{+1.4}_{-1.5}$     &  \phn2.9$^{+0.4}_{-0.4}$   &  \phn7.6$^{+1.0}_{-1.1}$   &  122/195\phn    & \phn8.7   \\
J1153+6753   &  1.5 & 1.75$^{+0.93}_{-0.45}$     &  0.06$^{+0.16}_{-0.06}$ &  10.9$^{+2.8}_{-1.9}$     &  \phn4.1$^{+1.0}_{-0.7}$   &  \phn9.7$^{+2.5}_{-1.7}$   &  128/224\phn    & \phn8.6   \\
J1336+5453   &  1.0 & 2.00                       &  0.30                   &  $<$ 4.9                  &  $<$ 1.5                   &  $<$ 3.5                   &  \phn$\cdots$/259\phn    & $\cdots$   \\
J1410+4145   &  1.5 & 1.62$^{+0.29}_{-0.20}$     &  0.33$^{+0.21}_{-0.14}$ &  25.1$^{+1.9}_{-4.9}$     &  \phn5.8$^{+0.4}_{-1.1}$   &  13.0$^{+1.0}_{-2.5}$      &  334/508\phn    & 14.8  \\
J1411+5736   &  1.2 & 1.57$^{+0.79}_{-0.57}$     &  0.14$^{+0.62}_{-0.14}$ &  \phn5.4$^{+0.7}_{-1.2}$  &  \phn1.6$^{+0.2}_{-0.3}$   &  \phn3.7$^{+0.5}_{-0.8}$   &  109/213\phn    & \phn7.5   \\
\enddata
\tablenotetext{a}{$kT$ was fixed to 2 keV to estimate
detection limits for undetected sources and J0133$-$1026.}
\tablenotetext{b}{Abundance assumes the solar photospheric values of \citet{AndersGrevesse1989}.  The value was fixed to 0.3 for fits with unconstrained abundance.}
\tablenotetext{c}{Absorbed model flux in the 0.5--2 keV band.  Upper limits are 3$\sigma$.}
\tablenotetext{d}{Unabsorbed model luminosity in the 0.5--2 keV band, rest frame.  Upper limits are 3$\sigma$.}
\tablenotetext{e}{Unabsorbed model luminosity in the 0.008--100 keV band, rest frame.  Upper limits are 3$\sigma$.}
\tablenotetext{f}{Counts in the spectral extraction region in the 0.5--2 keV band, observed frame.  Source counts are estimated from the spectral model.}
\tablenotetext{g}{Detection significance in units of $\sigma$.}
\end{deluxetable*}

\begin{deluxetable*}{cccccccccc}
\tabletypesize{\footnotesize}
\tablewidth{0pt}
\tablecaption{Spatial Fitting Results
\label{tab:spatial}}
\tablehead{
\colhead{FG} &
\colhead{kpc/arcmin} &
\colhead{$r_{extract}$} &
\colhead{$\beta$} &
\colhead{$r_{c}$} &
\colhead{$r_{500}$} &
\colhead{$M_{500}$} &
\colhead{ap.~cor.\tablenotemark{a}} &
\colhead{$L_{soft,500}$\tablenotemark{b}} &
\colhead{$L_{bol,500}$\tablenotemark{c}} \\
\colhead{} &
\colhead{} &
\colhead{(arcmin, kpc)} &
\colhead{} &
\colhead{(kpc)} &
\colhead{(kpc)} &
\colhead{(\eez{13} \msun)} &
\colhead{} &
\colhead{(\eez{42} cgs)} &
\colhead{(\eez{42} cgs)}
}
\startdata
J0133$-$1026 & 124 & 2.07 256 & \nodata                   & \nodata               & \nodata                &   \nodata                  &  \nodata              & \nodata                  & \nodata                      \\
J0815+3959   & 139 & 1.80 249 & 0.40                      & 25                    & 443$^{+59}_{-71\phn}$  &   \phn2.5$^{+1.1}_{-1.0}$  &  1.5$^{+0.1}_{-0.2}$  & \phn5.4$^{+1.1}_{-1.4}$  & 11.4$^{+2.3}_{-3.0\phn}$     \\
J0821+4051   & 135 & 1.85 250 & \nodata                   & \nodata               & \nodata                &   \nodata                  &  \nodata              & \nodata                  & \nodata                      \\
J0856+0553   & 105 & 2.12 223 & 0.32$^{+0.03}_{-0.03}$    & \phn7$^{+9\phn}_{-6}$ & 587$^{+111}_{-71}$     &   \phn5.7$^{+3.9}_{-1.8}$  &  2.8$^{+0.3}_{-0.2}$  & 13.7$^{+2.0}_{-1.7}$     & 35.9$^{+5.3}_{-4.6\phn}$     \\
J0906+0301   & 145 & 1.72 250 & \nodata                   & \nodata               & \nodata                &   \nodata                  &  \nodata              & \nodata                  & \nodata                      \\
J1007+3800   & 123 & 2.07 253 & 0.50$^{+0.09}_{-0.07}$    & 50$^{+19}_{-15}$      & 712$^{+103}_{-93}$     &   10.3$^{+5.1}_{-3.6}$     &  1.6$^{+0.3}_{-0.3}$  & 17.6$^{+3.4}_{-4.7}$     & 45.5$^{+8.7}_{-12.1}$        \\
J1017+0156   & 128 & 1.46 187 & 0.47$^{+0.12}_{-0.08}$    & 20$^{+15}_{-12}$      & 626$^{+157}_{-105}$    &   \phn7.0$^{+6.7}_{-3.0}$  &  1.7$^{+0.5}_{-0.5}$  & \phn3.2$^{+1.1}_{-1.1}$  & \phn7.8$^{+2.6}_{-2.6\phn}$  \\
J1039+3947   & 104 & 1.24 129 & 0.40                      & 25                    & 512$^{+129}_{-89}$     &   \phn3.8$^{+3.7}_{-1.7}$  &  3.0$^{+1.6}_{-1.0}$  & \phn3.2$^{+1.9}_{-1.3}$  & \phn7.4$^{+4.2}_{-3.0\phn}$  \\
J1045+0420   & 161 & 2.21 356 & 0.40                      & 25                    & 621$^{+102}_{-107}$    &   \phn6.8$^{+4.0}_{-2.9}$  &  1.5$^{+0.1}_{-0.2}$  & 25.7$^{+3.5}_{-5.1}$     & 65.3$^{+8.9}_{-13.0}$        \\
J1133+5920   & 142 & 1.20 171 & 0.40                      & 25                    & 496$^{+88}_{-82\phn}$  &   \phn3.5$^{+2.2}_{-1.5}$  &  2.3$^{+0.8}_{-0.7}$  & \phn7.7$^{+2.7}_{-2.8}$  & 17.0$^{+6.0}_{-6.2\phn}$     \\
J1136+0713   & 114 & 1.25 142 & 0.40                      & 25                    & 642$^{+169}_{-119}$    &   \phn7.5$^{+7.7}_{-3.5}$  &  3.2$^{+2.0}_{-1.2}$  & \phn9.5$^{+6.0}_{-3.9}$  & 24.6$^{+15.5}_{-10.1}$       \\
J1153+6753   & 127 & 1.37 174 & 0.40                      & 25                    & 522$^{+136}_{-104}$    &   \phn4.0$^{+4.1}_{-2.0}$  &  2.3$^{+0.7}_{-0.7}$  & \phn9.5$^{+3.7}_{-3.1}$  & 22.6$^{+8.9}_{-7.5\phn}$     \\
J1336+5453   & 118 & 2.12 250 & \nodata                   & \nodata               & \nodata                &   \nodata                  &  \nodata              & \nodata                  & \nodata                      \\
J1410+4145   & 105 & 2.03 213 & 0.39$^{+0.06}_{-0.04}$    & 25$^{+15}_{-11}$      & 496$^{+53}_{-41\phn}$  &   \phn3.5$^{+1.2}_{-0.8}$  &  2.0$^{+0.2}_{-0.3}$  & 11.5$^{+1.2}_{-2.7}$     & 25.9$^{+2.8}_{-6.1\phn}$     \\
J1411+5736   & 117 & 1.34 156 & 0.40                      & 25                    & 495$^{+124}_{-126}$    &   \phn3.4$^{+3.3}_{-2.0}$  &  2.5$^{+0.5}_{-0.7}$  & \phn4.0$^{+1.0}_{-1.5}$  & \phn9.3$^{+2.3}_{-3.4\phn}$  \\
\enddata
\tablenotetext{a}{Aperture correction factor to convert $L_X$ through the observed aperture to \rfh.}
\tablenotetext{b}{Unabsorbed model luminosity in the 0.5--2 keV band, rest frame, corrected to \rfh.}
\tablenotetext{c}{Unabsorbed model luminosity in the 0.008--100 keV band, rest frame, corrected to \rfh.}
\tablenotetext{}{}
\end{deluxetable*}

Of the 12 FGs detected in extended X-ray emission, four have sufficient
counts to constrain the $\beta$ model parameters.  These are four of the
five brightest targets, with $F \ge 20\eex{-14}$ \ergscms\ (0.5--2 keV).
The fifth bright target, J1045$+$0420, is morphologically irregular and
obviously not well-fit by a simple $\beta$ profile (see
Section~\ref{sect:indiv}).  The best-fit $\beta$ values range from 0.3 to
0.5, smaller than the value of 0.67 commonly found for clusters
\citep[e.g.,][]{JonesForman1999} but not unusually small for rich groups
and poor clusters of similar temperature in this redshift range
\citep{Willisetal2005,Jeltemaetal2006} or at $z \sim 0$
\citep{OsmondPonman2004}.  The values for the core radius \rc\ are all
comparatively small, ranging from 7 to 50 kpc, but consistent with the
previously cited results.  For the seven detections with unconstrained
$\beta$ model parameters, we assumed the average best-fit values of $\beta
= 0.4$ and $\rc = 25$ kpc to estimate the spatial extent of the X-ray
emission and the luminosity corrections.  Note that despite an X-ray
detection, J0133$-$1026 was excluded from the remaining analysis, since its
lack of a measured temperature rendered the spatial extrapolation too
uncertain.

Based on the best-fit $kT$, $\beta$, and $\rc$ with associated errors, we
estimated \rfh\ and \mfh\ for each group.  The mass within radius $r$ can
be given as \citep[e.g.,][]{ArnaudEvrard1999}:
\begin{equation}
M(<r) = 1.13\eex{14}~\beta~\frac{kT}{{\rm keV}}~\frac{r}{{\rm Mpc}}~\frac{(r/\rc)^2}{1+(r/\rc)^2}~\msun.
\label{eq:mass}
\end{equation}
Since $\rfh \gg \rc$, we can estimate \rfh\ with a simple analytic
approximation \citep[e.g.,][Eq.~17]{ArnaudEvrard1999}.  We chose to
iteratively solve Eq.~\ref{eq:mass} for \mfh\ and \rfh, using the
definition 
\begin{equation}
\rfh = \left[\frac{3\mfh}{4 \pi 500 \rho_{crit}(z)}\right]^{1/3}.
\end{equation}
The results are shown in Table~\ref{tab:spatial}; \rfh\ = 443--712 kpc
and \mfh\ = 0.3--1.0\eex{14} \msun\ for the sample, typical values for
groups and clusters in this temperature and redshift range
\citep{Willisetal2005,Jeltemaetal2006,Finoguenovetal2007,Jeltemaetal2009}.
These values are equivalent to what we derive from the analytic
approximation.

Aperture corrections were calculated to scale the observed luminosity to
that within \rfh, including the small ($< 2\%$) correction for excluded
point source regions.  The aperture corrections range from 1.5 to 3.0 with 
an average correction of 2.2.
While these are large corrections, the attendant errors take into
account the uncertainty in the $\beta$ model parameters.  The
corrections for the four groups with well-constrained $\beta$ models are
1.6--2.8, with errors of less than 30\%.  The aperture corrections and
corrected \rfh\ luminosities are listed in Table~\ref{tab:spatial}.

% XXX something about not being able to do morphology
To analyze the structure of the hot intra-cluster medium (ICM), researchers
typically employ power ratios (multipole moments of the X-ray surface
brightness) and centroid shifts that are sensitive to substructure and
irregular morphology \citep[e.g.,][]{Jeltemaetal2008}.  Power ratios
require thousands of counts and a well-constrained surface brightness out
to $\sim$ \rfh, therefore we cannot employ them with the current snapshot
data.  We attempted to calculate centroid shifts for the detected FGs,
however the results were inconclusive, with large errors driven by the low
counting statistics.  We present a qualitative analysis of the morphology,
specifically the fraction of disturbed and relaxed clusters, in
Section~\ref{sect:success}.

\begin{deluxetable}{cccccc}[H]
\tabletypesize{\footnotesize}
\tablewidth{0pt}
\tablecaption{BCG X-ray Emission
\label{tab:ptsrc}}
\tablehead{
\colhead{FG} &
\colhead{$\Gamma$\tablenotemark{a}} &
\colhead{flux\tablenotemark{b}} &
\colhead{\underline{BCG}\phn} &
\colhead{$L_{soft}$\tablenotemark{d}} &
\colhead{cts\tablenotemark{e}} \\
\colhead{} &
\colhead{} &
\colhead{(\eez{-14} cgs)} &
\colhead{IGM\tablenotemark{c}} &
\colhead{(\eez{42} cgs)} &
\colhead{}
}
\startdata
J0133$-$1026 & 1.8$^{+0.2}_{-0.4}$ & 1.03$^{+0.40}_{-0.41}$ & 0.30 & 0.33$^{+0.13}_{-0.13}$ & 35 \\
J1007$+$3800 & 1.7                 & 1.38$^{+0.58}_{-0.55}$ & 0.04 & 0.45$^{+0.19}_{-0.18}$ & 23 \\
J1017$+$0156 & 1.7                 & 0.42$^{+0.19}_{-0.21}$ & 0.09 & 0.17$^{+0.08}_{-0.08}$ & 15 \\
J1039$+$3947 & 1.7                 & 0.94$^{+0.23}_{-0.23}$ & 0.20 & 0.21$^{+0.05}_{-0.05}$ & 17 \\
J1045$+$0420 & 1.7                 & $<$0.22                & $<$0.01 & $<$0.14 & \phn7 \\
J1153$+$6753 & 1.7                 & 1.50$^{+0.50}_{-0.50}$ & 0.14 & 0.55$^{+0.18}_{-0.18}$ & 20 \\
J1410$+$4145 & 1.7                 & 1.67$^{+0.61}_{-0.59}$ & 0.07 & 0.38$^{+0.14}_{-0.13}$ & 30 \\
J1411$+$5736 & 1.7                 & 1.25$^{+0.50}_{-0.50}$ & 0.23 & 0.37$^{+0.15}_{-0.15}$ & 22 \\
\enddata
\footnotesize
\tablenotetext{a}{$\Gamma$ is the power law index, fixed to 1.7 for all but one source.}
\tablenotetext{b}{Absorbed model flux in the 0.5--2 keV band.}
\tablenotetext{c}{Ratio of the BCG 0.5-2 keV flux to the detected diffuse flux from \\\hspace*{1em}Table~\ref{tab:spectral}.}
\tablenotetext{d}{Unabsorbed model luminosity in the 0.5--2 keV band, rest frame.}
\tablenotetext{e}{Counts in BCG region, 0.3--7 keV.  No background has been \\\hspace*{1em}subtracted.}
\end{deluxetable}

\subsection{BCG X-ray Emission}
\label{sect:ptsrc}

\begin{figure*}
\centering
\begin{minipage}[t]{.32\linewidth}
\centering
\includegraphics[width=\linewidth]{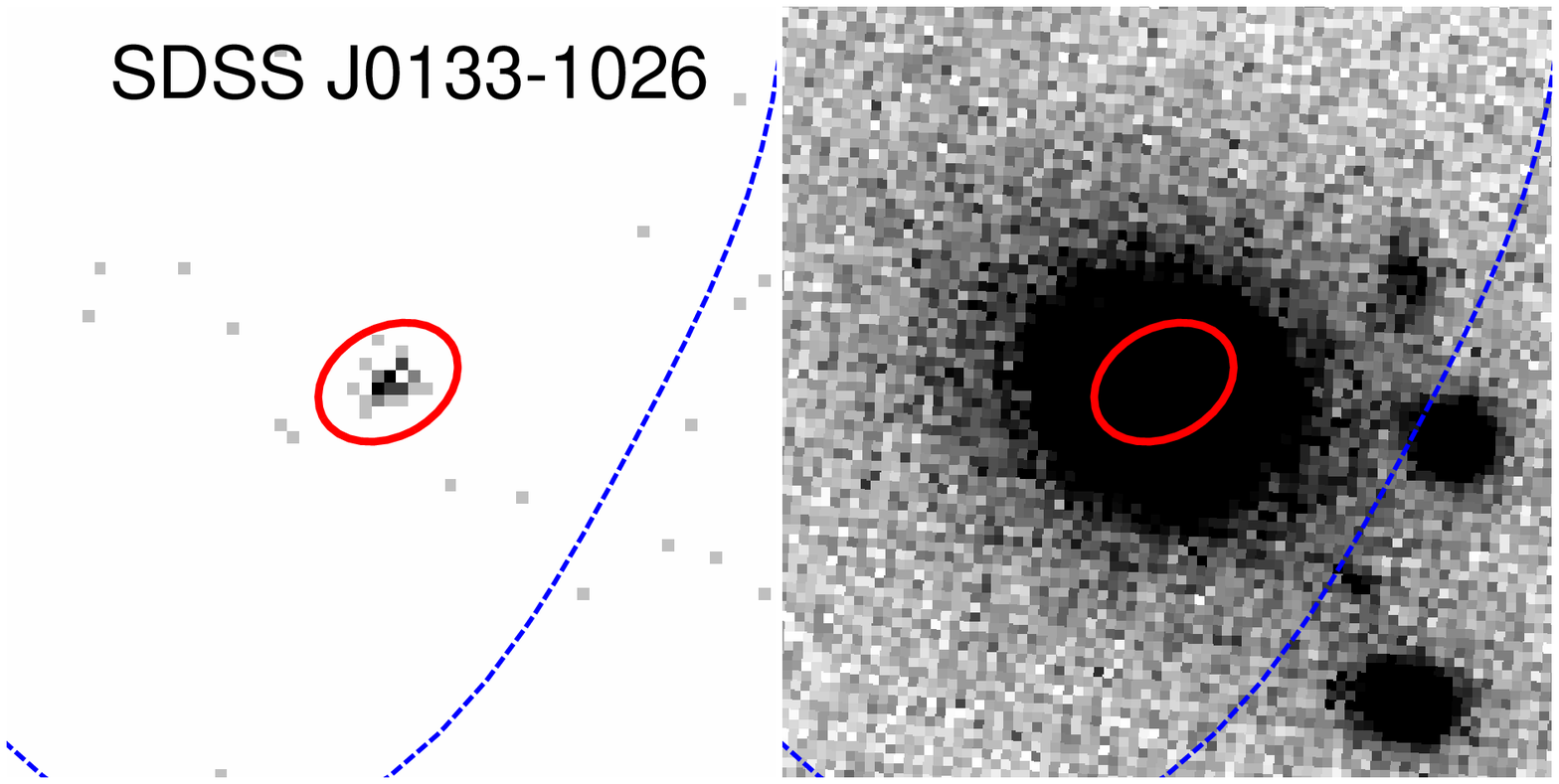}
\end{minipage}
\begin{minipage}[t]{.32\linewidth}
\centering
\includegraphics[width=\linewidth]{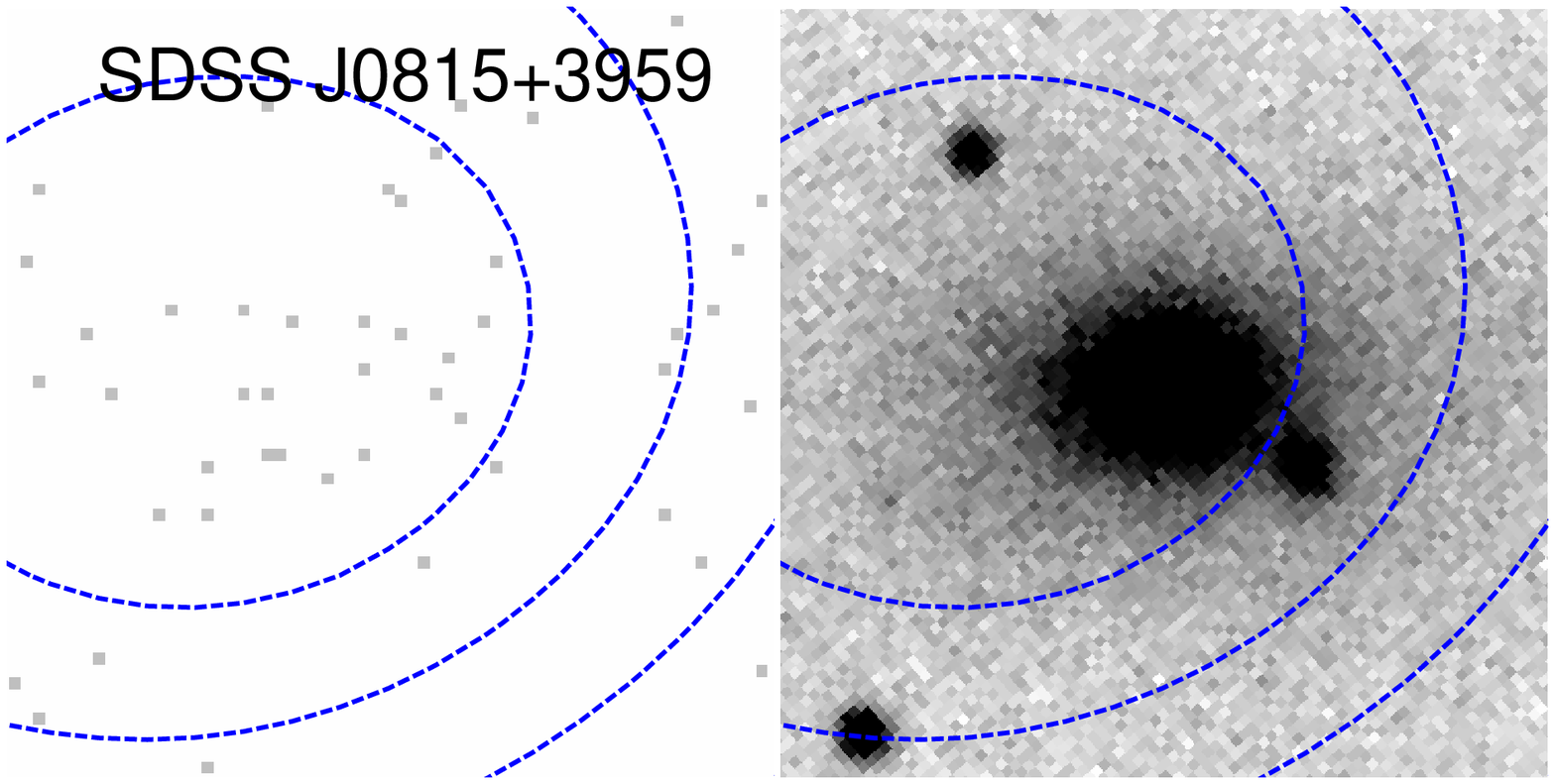}
\end{minipage}
\begin{minipage}[t]{.32\linewidth}
\centering
\includegraphics[width=\linewidth]{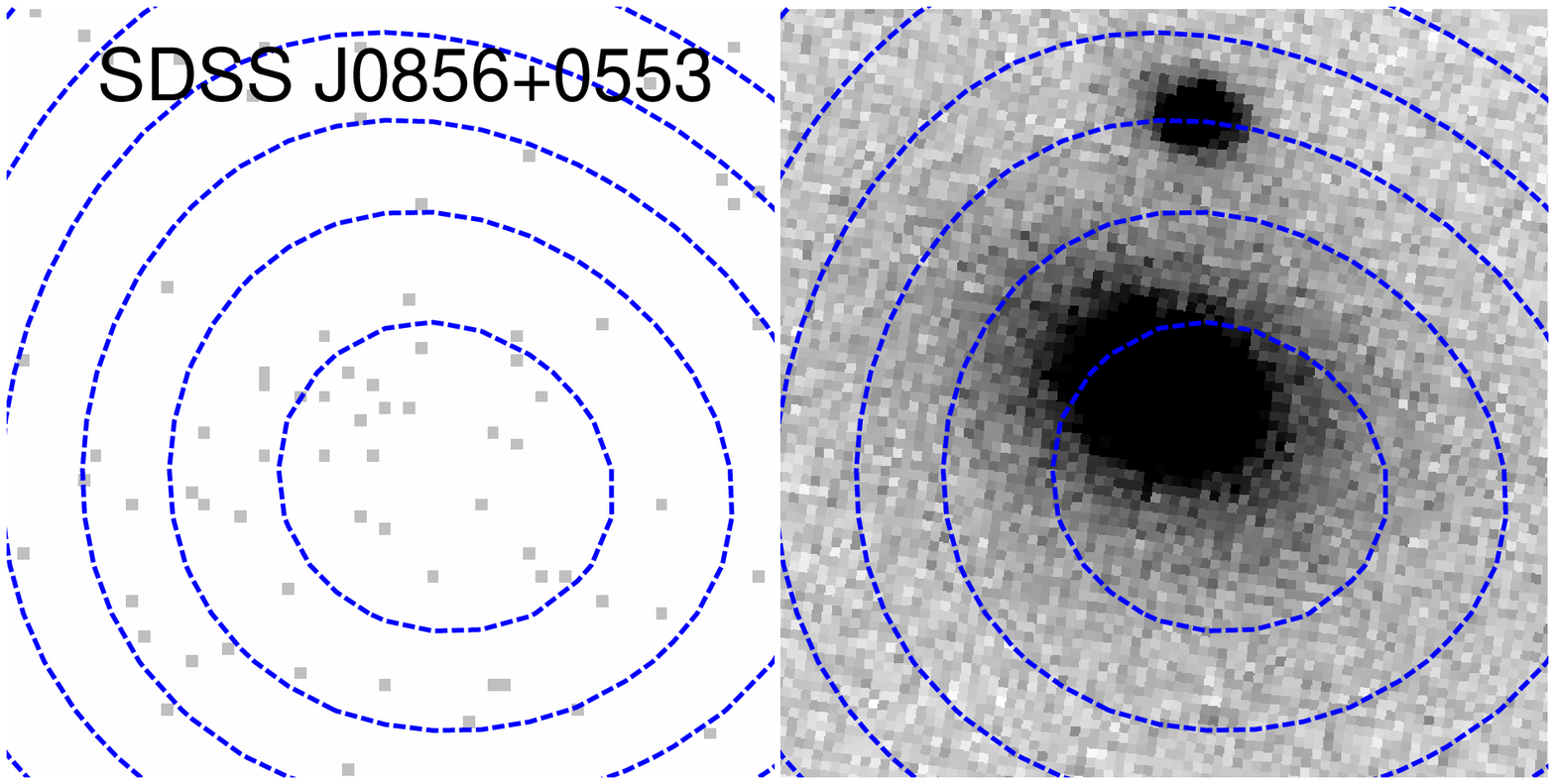}
\end{minipage}
\\
\begin{minipage}[t]{.32\linewidth}
\centering
\includegraphics[width=\linewidth]{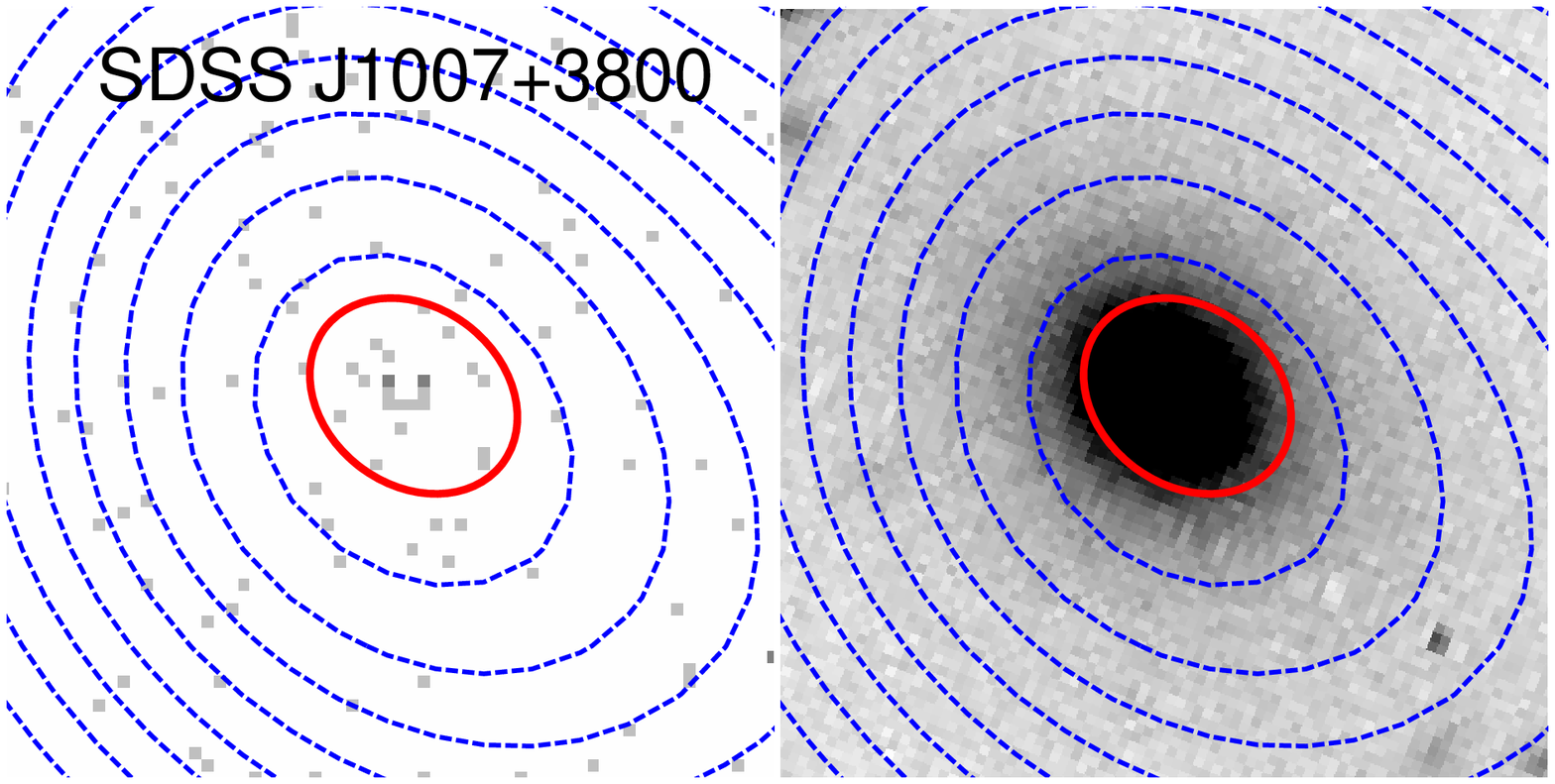}
\end{minipage}
\begin{minipage}[t]{.32\linewidth}
\centering
\includegraphics[width=\linewidth]{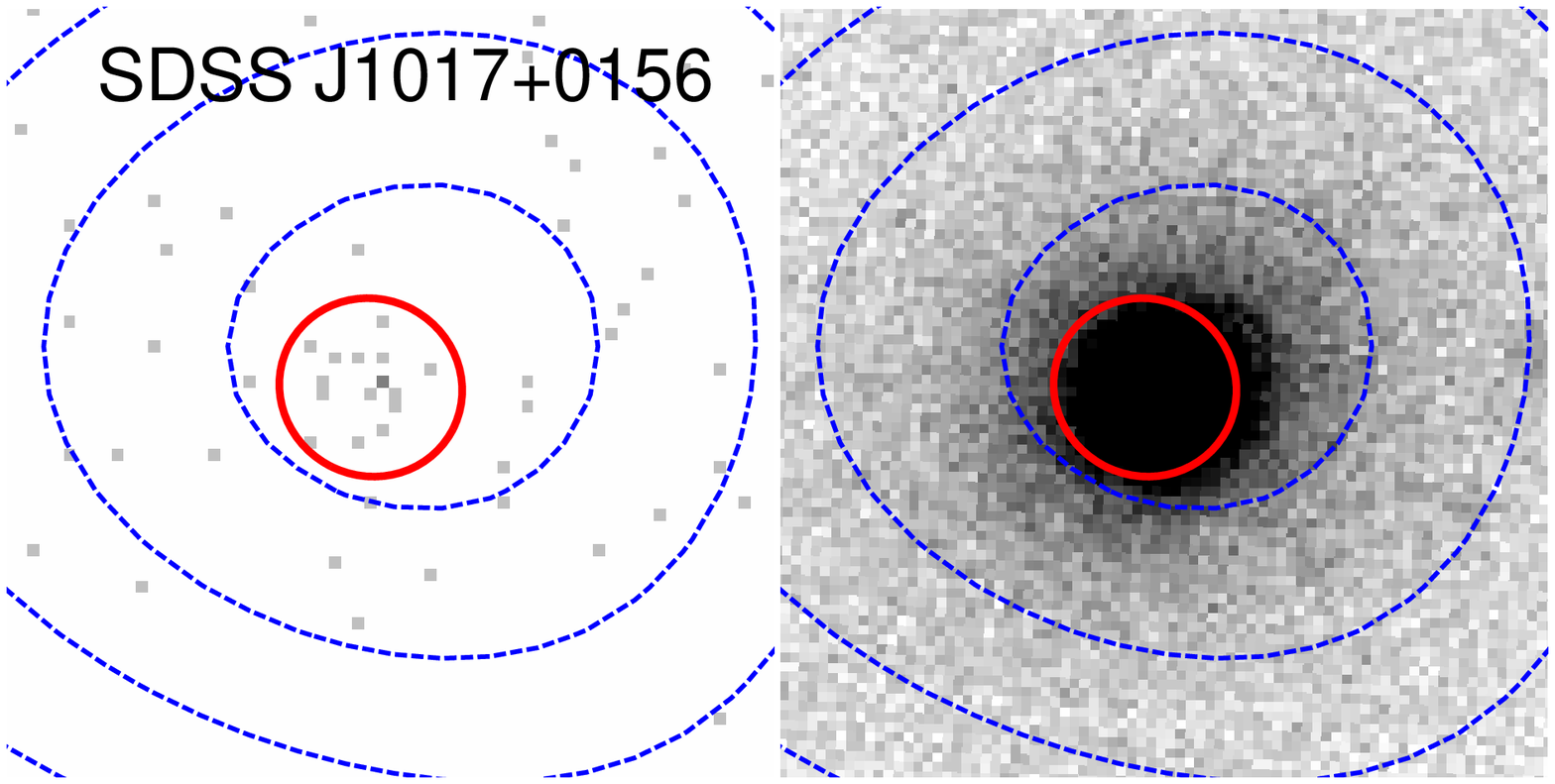}
\end{minipage}
\begin{minipage}[t]{.32\linewidth}
\centering
\includegraphics[width=\linewidth]{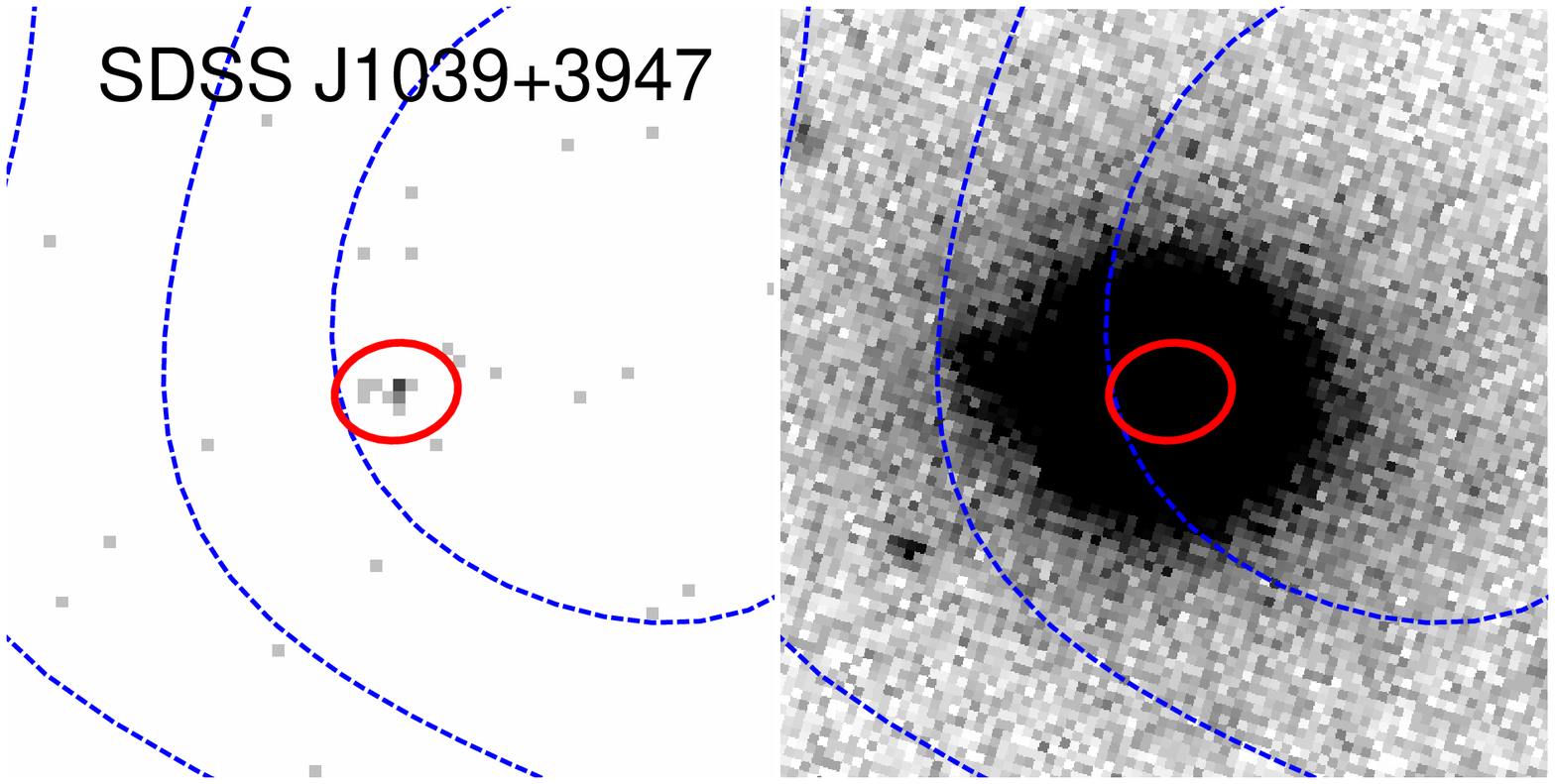}
\end{minipage}
\\
\begin{minipage}[t]{.32\linewidth}
\centering
\includegraphics[width=\linewidth]{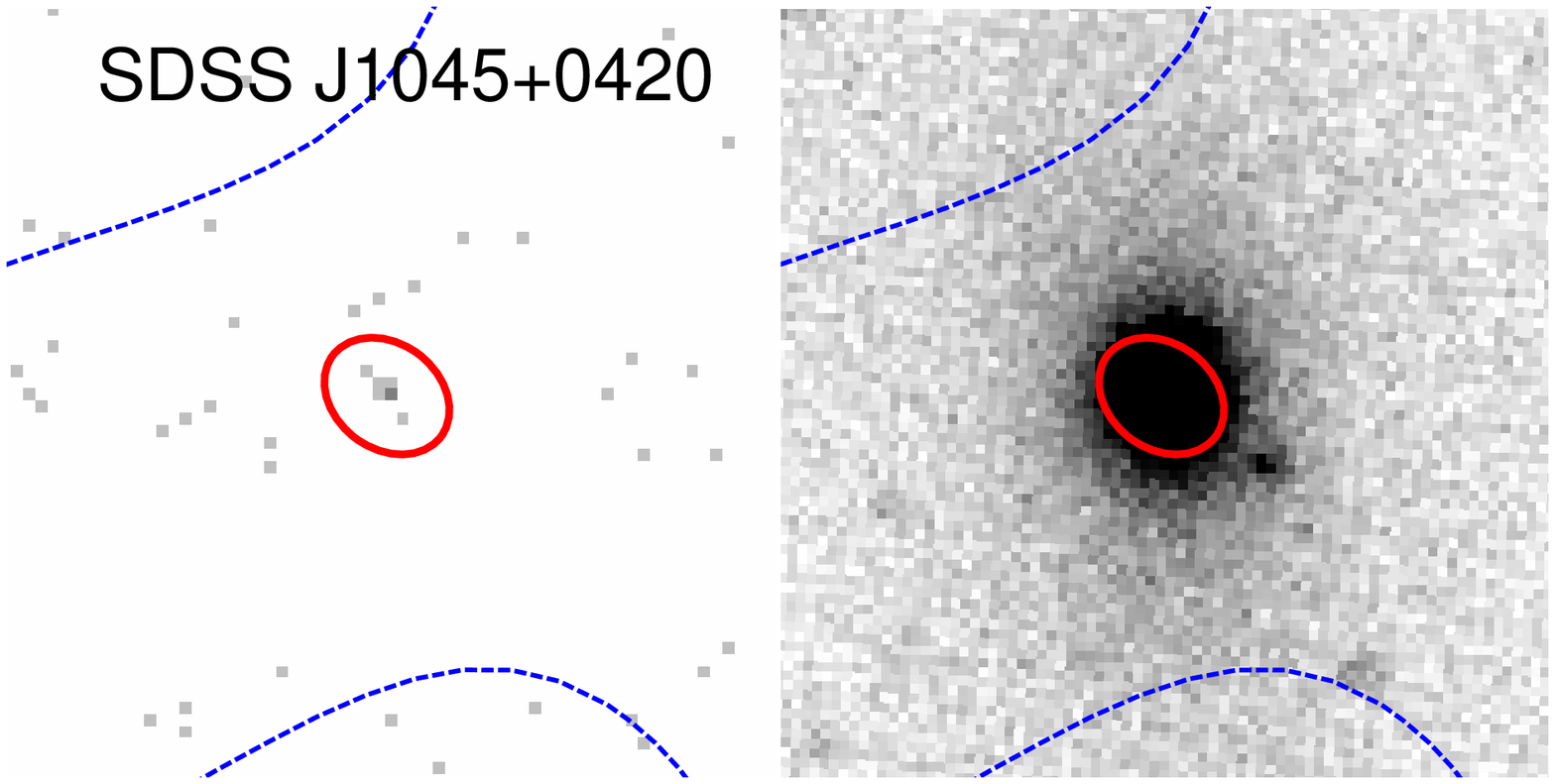}
\end{minipage}
\begin{minipage}[t]{.32\linewidth}
\centering
\includegraphics[width=\linewidth]{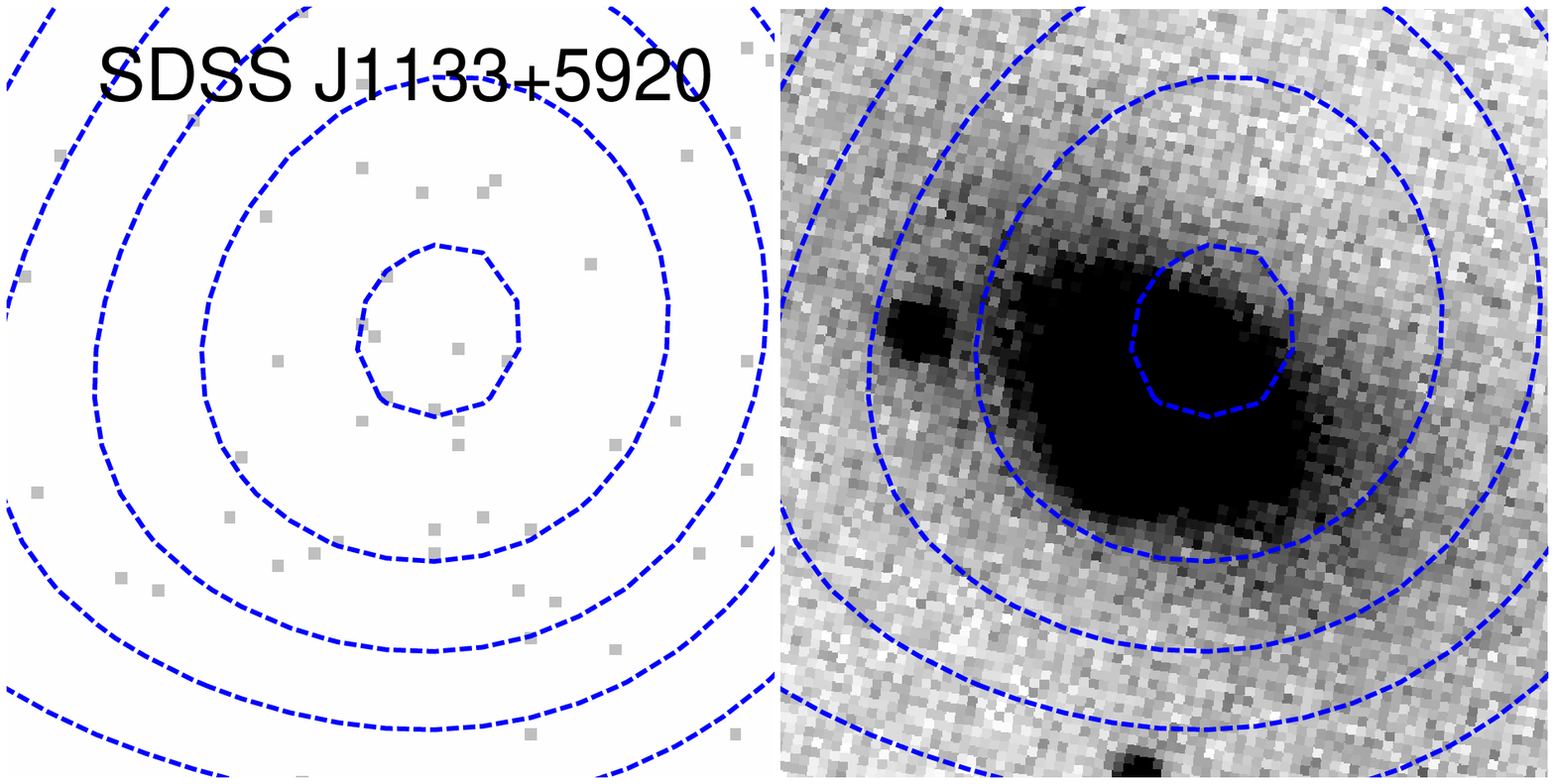}
\end{minipage}
\begin{minipage}[t]{.32\linewidth}
\centering
\includegraphics[width=\linewidth]{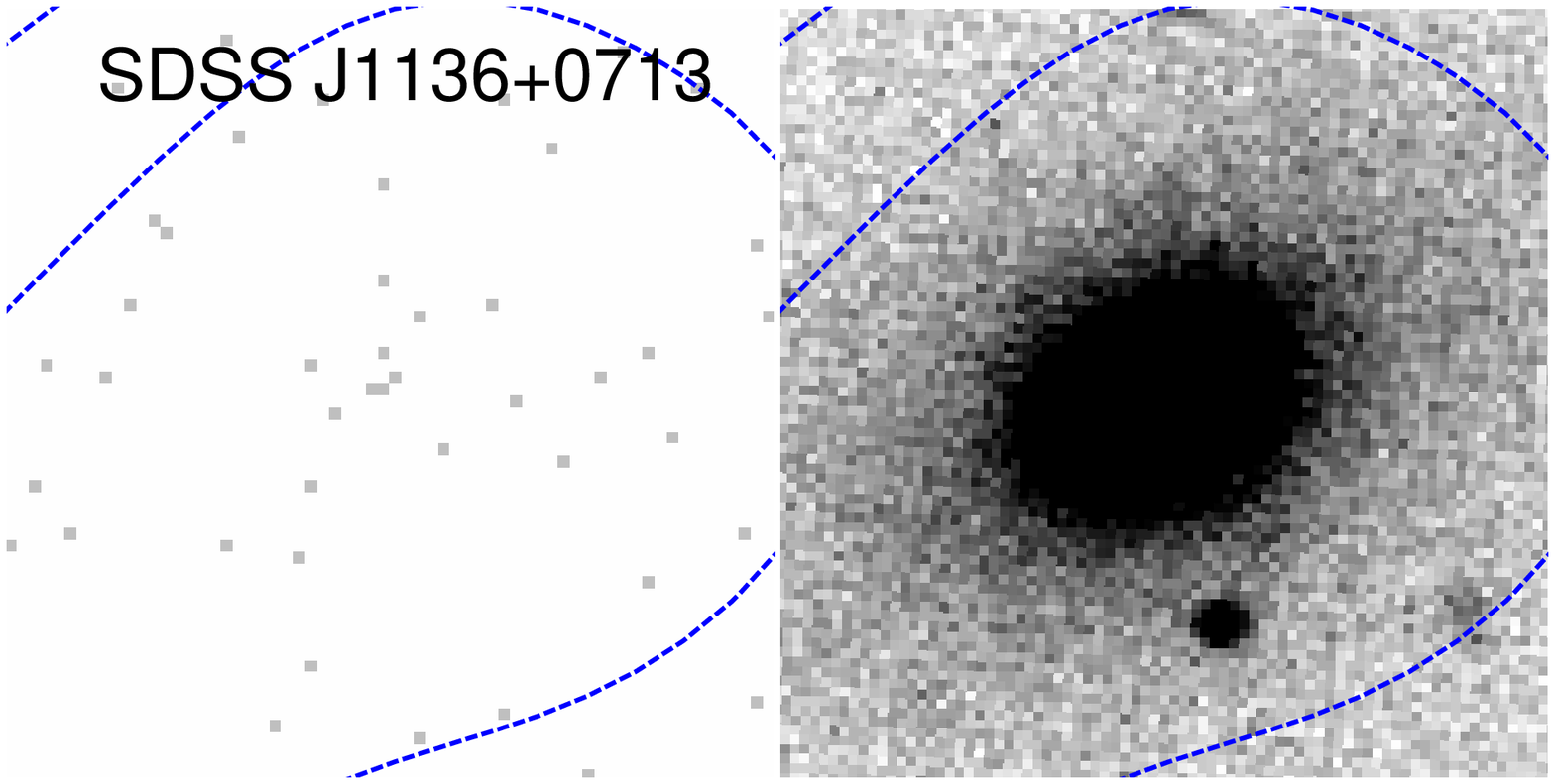}
\end{minipage}
\\
\begin{minipage}[t]{.32\linewidth}
\centering
\includegraphics[width=\linewidth]{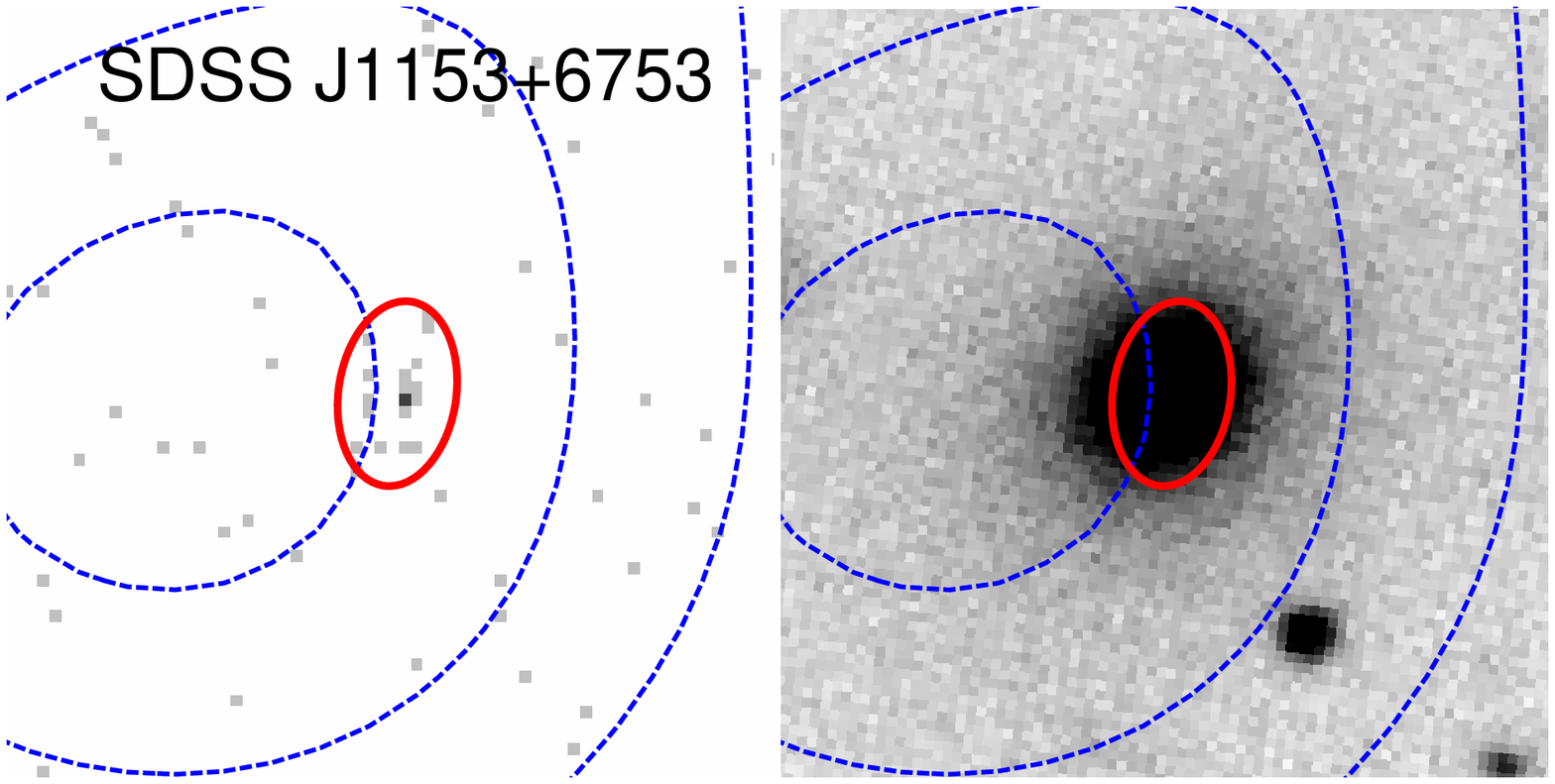}
\end{minipage}
\begin{minipage}[t]{.32\linewidth}
\centering
\includegraphics[width=\linewidth]{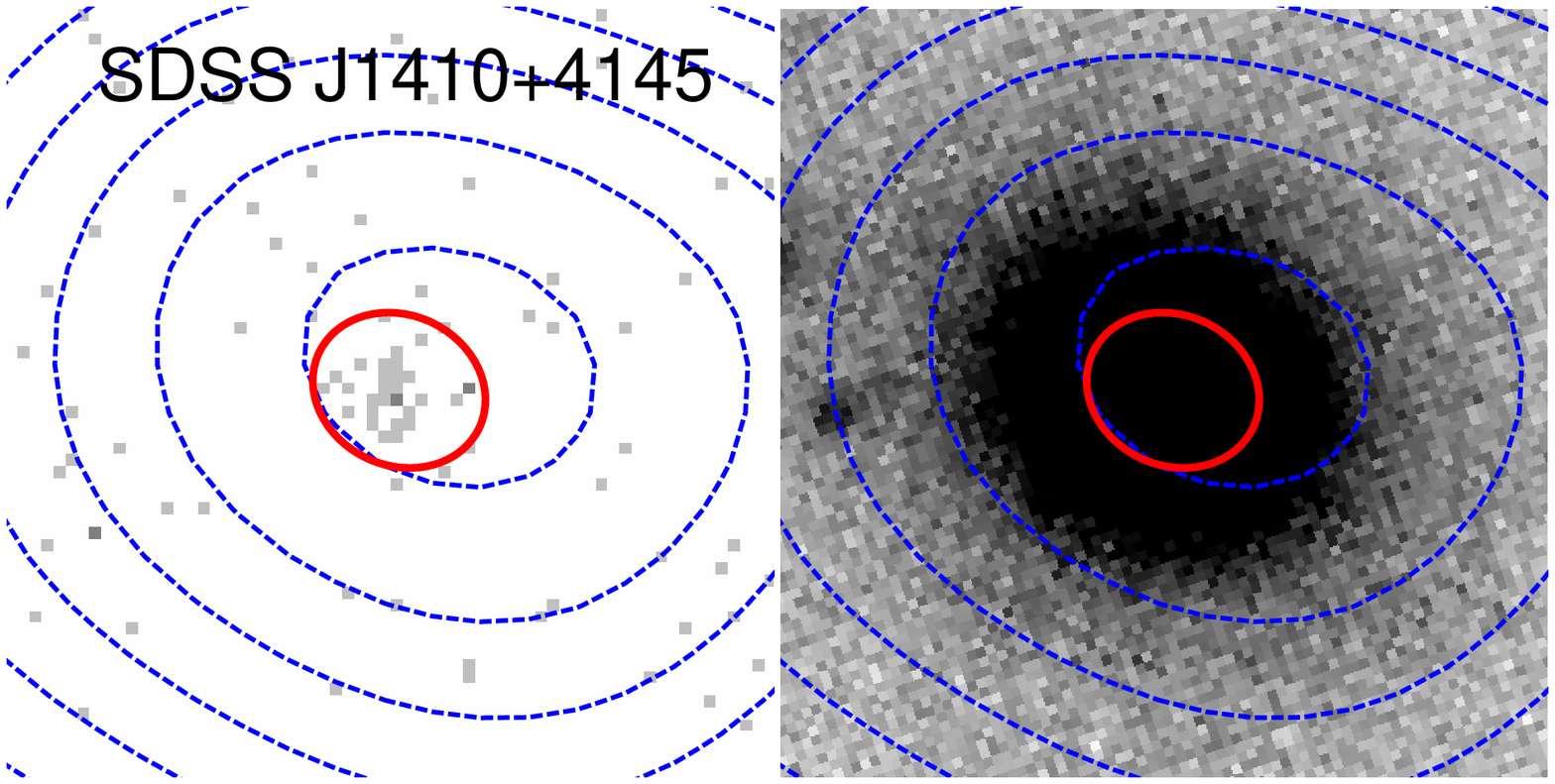}
\end{minipage}
\begin{minipage}[t]{.32\linewidth}
\centering
\includegraphics[width=\linewidth]{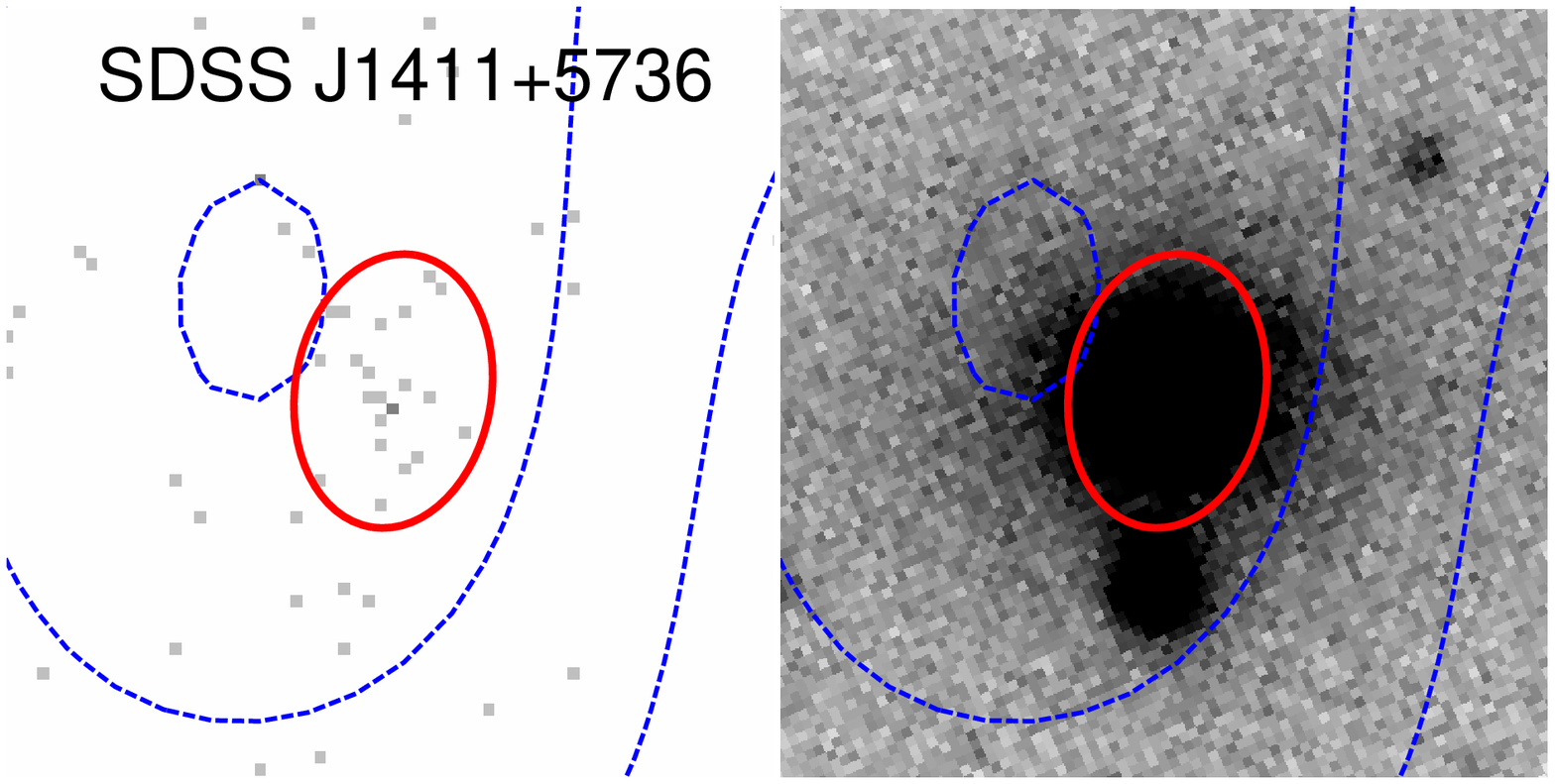}
\end{minipage}
\caption{\chandra\ 0.3--7 keV counts image (left) and SDSS $r$-band image
(right) for each FG, centered on the BCG.  Blue dashed contours show the
same point-source-excluded, smoothed emission as in
Figure~\ref{fig:smimages}.  Red ellipses identify BCG emission that has
been excluded from the spectral and spatial analysis of the IGM emission
for eight of the FGs, and analyzed separately in Section~\ref{sect:ptsrc}.
Images are 30\arcsec\ on a side.}
\label{fig:ptsrc}
\end{figure*}

To reduce the possibility of contamination from AGN or diffuse emission
in the BCG, we excluded sources found by {\tt wavdetect} in
Section~\ref{sect:data} that fall within the $r$-band optical extent of the
BCG.  As the images in Figure~\ref{fig:ptsrc} show, 8 of the 12 BCGs
possess a detected central source with emission in the 0.3--7 keV band,
ranging from 7 to 35 counts in the regions outlined by red ellipses.
Several of these sources appear to be point-like AGN, but all 12 FGs have
some extended component as well, either from the BCG, the core of the IGM,
or some combination thereof.

We estimated the contribution of these detected sources to the total X-ray
flux by extracting a spectrum for each source and fitting with XSPEC.
We assumed a spectral model typical of an AGN, an absorbed power law
with fixed power law index $\Gamma = 1.7$, and used the fixed $N_H$ and redshift
values from Table~\ref{tab:spectral}.  The power law normalization was
allowed to vary, which provided a confidence interval for the source flux.
For one of the targets, the source in J0133$-$1026 with 35 counts, we
allowed $\Gamma$ to vary and obtained a good constraint of $\Gamma =
1.8^{+0.2}_{-0.4}$.  The 0.5--2 keV flux estimates (see
Table~\ref{tab:ptsrc}) range from a few percent to 30 percent of the flux
from the diffuse IGM emission.  

One or more of the detected central X-ray sources could be thermal in
origin, along the lines of the embedded coronae identified in cluster
galaxies by \citet{Sunetal2007}.  This is especially true of the more
extended emission sources, such as those in SDSS J1007$+$3800, SDSS
J1410$+$4145, and SDSS J1411+5736.  While the spectral model used here is
inappropriate for thermal BCG X-ray emission, it nevertheless provides a
reasonable estimate of the flux in the observed band for planning future
deep X-ray observations.  Due to the paucity of counts from these sources,
a detailed spectral analysis is not possible with the current data.  FGs
with notable BCG X-ray flux are discussed in more detail in
Section~\ref{sect:indiv}.  We emphasize that we have excluded the BCG
emission in our spectral (Section~\ref{sect:spectral}) and spatial
(Section~\ref{sect:spatial}) analysis for the 8 clusters in which we detect
it.  Therefore the total X-ray flux for an instrument which is unable to
resolve and exclude these central sources would be the sum of the values
listed in Tables~\ref{tab:spectral} and \ref{tab:ptsrc}.

\section{RESULTS AND DISCUSSION}
\label{sect:results}

%\subsection{Success of the Method:  Are They \\``Real'' Fossil Groups?}
\subsection{\texorpdfstring{Success of the Method:  Are They \\``Real'' Fossil Groups?}{Success of the Method:  Are They ``Real'' Fossil Groups?}}
\label{sect:success}

The primary goal of this work, to optically identify a sizable sample of
fossil groups, has met with great success.  Out of a sample of 15
candidates, we have confirmed 12 gravitationally bound systems through
their IGM X-ray emission.  One additional target, SDSS J0906+0301, is
undetected in the short \chandra\ exposure but is clearly a bound system
from follow-up optical spectroscopy, with a velocity dispersion of $\sigma
= 506\pm72$ \kps\ based on 25 member redshifts \citep{Proctoretal2011}.  This
87\% success rate demonstrates the value of the maxBCG survey for selecting
FGs; moreover, since these galaxy-poor systems are the most difficult to
identify optically, we expect a very high success rate for this method in
selecting clusters and groups in general.  However, we note two caveats
that must be addressed.

First, our original FG criteria were based on a mass-scaling relation from
SDSS systems \citep{Johnstonetal2007}.  From the current analysis
and the results of \citet{Proctoretal2011}, the systems in our sample are
very massive for their richness, and in fact they are underluminous in the
optical (including the BCG luminosity) by about a factor of three.  Simply
scaling for the observed richness could dramatically underestimate \rth,
which in turn could exclude bright galaxies from our $\Delta_i$ magnitude
difference criterion and render our systems non-fossil groups.  To address
this, we have calculated new values for \rth\ based on the X-ray results
presented here.  In addition to the the $\beta$ model approach with which
we estimated \rfh, we adopt the scaling relation of
\citet{HelsdonPonman2003},
\begin{equation}
r_{200,kT} = 1.14~\left[\frac{kT}{\rm keV}\right]^{1/2}~h^{-1}_{50}~E(z)^{-1}~{\rm Mpc},
\label{eq:r200}
\end{equation}
where $E(z) = H(z)/H_0 = [\Omega_m\,(1+z)^3 + \Omega_{\Lambda}]^{1/2}$ for
a $\Lambda$CDM universe \citep[e.g.,][]{Maughanetal2006}.  These values and
the $\beta$ model extrapolation values (denoted $r_{200,\beta}$) are
presented in Table~\ref{tab:r200}, along with the magnitude difference
$\Delta_i$ determined for each \rth\ and its 1$\sigma$ spread of values.
The \rth\ values originally used for the maxBCG optical selection are
hereafter denoted $r_{200,mB}$.

\begin{deluxetable*}{ccccr@{ (}lcr@{ (}lc}
\tabletypesize{\footnotesize}
\tablewidth{0pt}
\tablecaption{Comparison of \rth\ Estimates and Resulting $\Delta_i$
\label{tab:r200}}
\tablehead{
\colhead{FG} &
\colhead{$N_{200}$\tablenotemark{a}} &
\colhead{$r_{200,mB}$\tablenotemark{b}} &
\colhead{$\Delta_i$\tablenotemark{c}} &
\multicolumn{2}{c}{$r_{200,\beta}$\tablenotemark{d}} &
\colhead{$\Delta_i$\tablenotemark{c}} &
\multicolumn{2}{c}{$r_{200,kT}$\tablenotemark{e}} &
\colhead{$\Delta_i$\tablenotemark{c}} \\
\colhead{} &
\colhead{} &
\colhead{(kpc)} &
\colhead{(mag)} &
\multicolumn{2}{c}{(kpc)} &
\colhead{(mag)} &
\multicolumn{2}{c}{(kpc)} &
\colhead{(mag)} 
}
\startdata
J0815$+$3959 & 12 & 738 & 3.3 & \phn700 &589,795)  & 3.3 (3.3,3.1) & \phn863 &799,924)   & 3.1 (3.1,3.1) \\
J0856$+$0553 & 16 & 833 & 2.3 & \phn928 &817,1103) & 2.3 (2.3,2.3) &    1291 &1156,1553) & 1.7 (2.3,1.7) \\
J1007$+$3800 & 24 & 988 & 2.4 &    1128 &981,1291) & 2.2 (2.4,2.2) &    1252 &1123,1403) & 2.2 (2.2,1.6) \\
J1017$+$0156 & 12 & 738 & 2.3 & \phn991 &825,1238) & 1.8 (1.8,1.8) &    1130 &983,1413)  & 1.8 (1.8,1.8) \\
J1039$+$3947 & 14 & 788 & 2.5 & \phn809 &669,1016) & 2.5 (2.5,1.4) &    1014 &916,1273)  & 1.4 (2.5,1.4) \\
J1045$+$0420 & 13 & 764 & 2.1 & \phn981 &812,1144) & 2.1 (2.1,2.1) &    1195 &1082,1350) & 2.1 (2.1,1.2) \\
J1133$+$5920 & 13 & 764 & 2.3 & \phn784 &655,925)  & 2.3 (2.3,2.3) & \phn964 &882,1108)  & 2.3 (2.3,2.3) \\
J1136$+$0713 & 17 & 855 & 2.3 &    1016 &827,1284) & 0.5 (2.3,0.5) &    1266 &1123,1608) & 0.5 (0.5,0.5) \\
J1153$+$6753 & 17 & 855 & 2.2 & \phn825 &661,1041) & 2.2 (2.2,2.2) &    1023 &892,1295)  & 2.2 (2.2,1.2) \\
J1410$+$4145 & 21 & 934 & 2.3 & \phn784 &720,869)  & 2.3 (2.3,2.3) & \phn997 &937,1084)  & 2.3 (2.3,2.3) \\
J1411$+$5736 & 16 & 833 & 2.2 & \phn783 &584,978)  & 2.2 (2.2,2.2) & \phn974 &797,1221)  & 2.2 (2.2,1.6) \\
\enddata
\footnotesize
\tablenotetext{a}{$N_{200}$ is defined in Table~\ref{tab:obs}.}
\tablenotetext{b}{\rth\ estimate from \citet{Johnstonetal2007} used in our maxBCG optical selection, as discussed in Section~\ref{sect:opt}.}
\tablenotetext{c}{$\Delta_i$ is defined in Table~\ref{tab:obs}.  Each column is determined from the \rth\ preceding it, with values in parentheses from the 1$\sigma$ extrema of \rth.}
\tablenotetext{d}{\rth\ estimate from extrapolating our $\beta$ model fit, with 1$\sigma$ extrema in parentheses.}
\tablenotetext{e}{\rth\ estimate from \citet{HelsdonPonman2003} (also see Eq.~\ref{eq:r200}), with 1$\sigma$ extrema in parentheses.}
\end{deluxetable*}

The original $r_{200,mB}$ values are consistent within the errors of the
$\beta$ model results for all but two FGs.  For J1410$+$4145, $r_{200,mB}$
was an overestimate, so no additional bright galaxies are included using
the $r_{200,\beta}$ value.  For J1017$+$0156, $r_{200,mB}$ was an
underestimate, and one bright galaxy falls within $r_{200,\beta}$ such that
$\Delta_i = 1.8$ mag.  This barely fails the magnitude difference criterion
of $\Delta_i \ge 2$ mag; in the $r$-band, the difference is only 1.9 mag
\citep{Proctoretal2011}.  Two other FGs (J1039$+$3947 and J1136$+$0713) are
consistent with the $\Delta_i$ criterion at the lower bound of
$r_{200,\beta}$, but inconsistent at its upper bound.

\begin{figure*}
\centering
\includegraphics[height=.9\linewidth,angle=270]{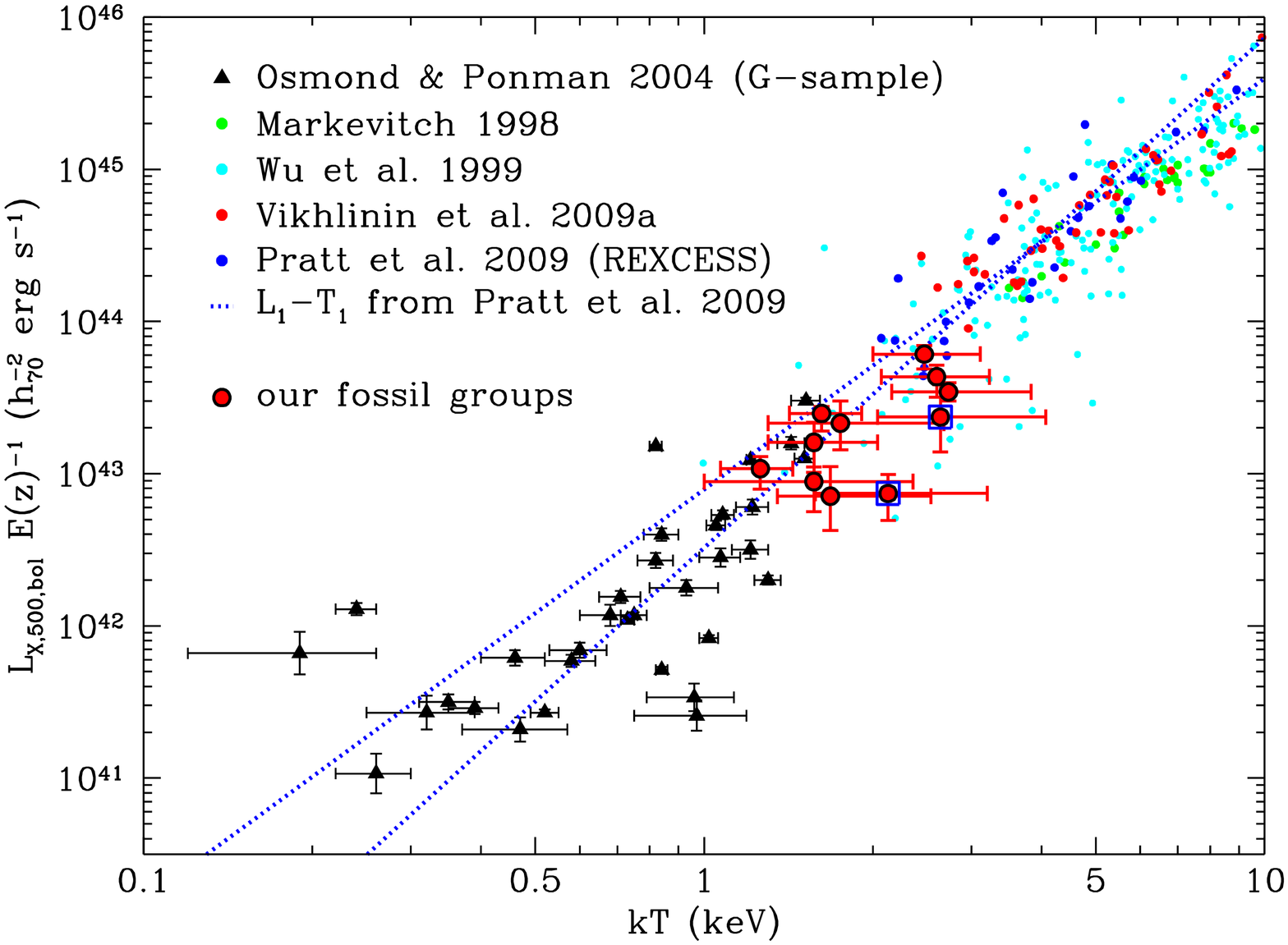}
\caption{$L_X$--$T_X$ relation for low-$z$ groups and clusters, with the
results overplotted for our 11 FGs with measured $T_X$.  All $L_X$ values
have been scaled to $h = 0.7$ and are measured within or corrected to \rfh\
except the \citet{Wuetal1999} values, which are within 1 Mpc.  The
values from \citet{Vikhlininetal2009a} are for the low-$z$ cluster sample,
and spectral temperatures exclude the inner 0.15\,\rfh.  The points marked
with blue boxes are discussed in Section~\ref{sect:success}.}
\label{fig:lt}
\end{figure*}

The $r_{200,kT}$ values are systematically $\sim 20\%$ higher than
$r_{200,mB}$ and $r_{200,\beta}$, however, the confidence intervals overlap
for 8 out of the 11 FGs.  Only the uncertainty in the X-ray $kT$
measurement has been propagated to the $r_{200,kT}$ confidence estimates,
therefore any intrinsic scatter in the Eq.~\ref{eq:r200} scaling relation
would increase the uncertainty.  With the current confidence intervals, two
FGs (J1017$+$0156 and J1136$+$0713) fail the $\Delta_i$ criterion for all
$\pm 1\sigma$ values of $r_{200,kT}$.  These systems are noted in
Figure~\ref{fig:lt} with blue squares.  A total of 8 FGs fail that
criterion for the extreme $+1\sigma$ value, but nevertheless agree with the
criterion within the uncertainty.  \citet{Proctoretal2011} reach identical
conclusions for the subset of 5 of these FGs for which they have $r$-band
photometry.

The second caveat concerns our X-ray surface brightness extrapolation.  The 
$\beta$ and core radii we have used are somewhat smaller than what is
typically found for clusters \citep[$\beta \sim 0.67$, $r_c \sim 100$--200
kpc;][]{JonesForman1999}.  Indeed, our extraction radii are not much
larger than the typical $r_c$ value, and it is possible the low $\beta$
values result from fitting a fairly flat core of emission.  For $\beta =
0.67$ and our assumed $\rc = 25$ kpc, the luminosity aperture corrections
to \rfh\ would be about 50\% lower than what is listed in
Table~\ref{tab:spatial}.  For $\rc = 100$ kpc, the aperture corrections are
25\% lower, and for $\rc = 200$ kpc, they are 20\% higher.  These
systematic uncertainties are within the statistical errors quoted for the
aperture corrections.

In summary, the $\Delta_i$ magnitude difference criterion is satisfied for
the majority of our FGs with measured temperature, insofar as our knowledge
of $\rth$ is correct, and therefore we conclude that these systems are real
fossil groups.  
Deeper X-ray observations will more precisely constrain the $kT$ and
surface brightness profiles to allow more accurate hydrostatic mass
determinations.  Regardless of the details of whether these systems
\textit{strictly\/} meet the empirical FG definition, the results presented
by \citet{Proctoretal2011} clearly demonstrate that a subset of our FGs
are different from other systems of similar mass.

As noted in Section~\ref{sect:opt}, the exclusion of candidates with known
central radio sources could bias our sample.  In particular, galaxy
clusters with strong cool cores are very likely to have radio AGN
\citep{Sunetal2009,Mittaletal2009}.  Thus to first appearances we have
selected against strong cool core FGs.  In fact, our screening criterion
turns out to be insufficient to remove all radio sources, as 4 of the 12
confirmed FGs have radio lobes that are outside of the 3\arcsec\ radius cut
that we imposed (see Section~\ref{sect:indiv}).  
Thus while some strong cool core FGs may have escaped our sample, it
remains useful for future radio studies, and we expect little bias in 
$kT$ given the large measurement uncertainties in the current analysis.  A
deeper \xmm\ study will allow removal of the core emission when assembling
scaling relations, a technique that greatly reduces scatter and does not
produce any apparent bias \citep{Prattetal2009,Vikhlininetal2009a}.

Fossil groups typically have relaxed X-ray morphology, and the
morphological differences in our sample are at first glance somewhat
puzzling.  As pointed out in Section~\ref{sect:spatial}, the modest
signal-to-noise of our snapshot data precludes a detailed analysis of the
structure of each FG.  From a qualitative standpoint, we conclude that much
of the appearance of disturbed morphology is due to lack of photons; this
is likely the case for J0133$-$1026, J0815$+$3959, J1039$+$3947, and
J1411$+$5736, and little can be said about the relaxed state of these four
FGs.  Five of the FGs are qualitatively relaxed, having bright core regions
and elliptical isophotes excluding the outer few contour levels (see
Figure~\ref{fig:smimages}); these are J0856$+$0553, J1017$+$0156,
J1133$+$5920, J1153$+$6753, and J1410$+$4145.  Of the three remaining FGs,
J1045$+$0420 is clearly disturbed, J1007$+$3800 has irregular isophotes
that are compressed on the east compared to the west, and J1136$+$0713 has
broad but low-level irregular emission.  Therefore, we can qualitatively
say that between 10\% (1 out of 8) and 40\% (3 out of 8) of the FGs are
apparently not relaxed.

These estimates for disturbed fraction are a bit lower than what is seen in
observations and simulations of normal systems.  \citet{Bohringeretal2010}
derive power ratios and centroid shifts for 31 clusters from the
Representative \xmm\ Cluster Structure Survey (REXCESS) and identify two
discrete samples in these parameter distributions, with $\sim$ 40\% of the
clusters identified as disturbed systems.  This is in line with previous
estimates of the disturbed fraction \citep[e.g.,][]{JonesForman1999}.
\citet{Jeltemaetal2008} perform simulations of clusters and measure the
same structure observables, and find that $\sim$ 30\% of the systems would
appear disturbed in \chandra-quality X-ray images.  Obviously the exact
fraction of disturbed and relaxed clusters depends on how these classes are
observationally defined and what metrics are used.  From a qualitative
view, the fraction of relaxed FGs in our sample is similar to what is
seen and predicted in normal systems, and in fact could be significantly
higher than that, more consistent with the expectation for FGs.

\subsection{Fossil Group Scaling Relations}
\label{sect:scaling}

With the scaled luminosities, and with the caveats outlined in
Section~\ref{sect:success} in mind, we are able to place the FGs on a
$L_X$--$T_X$ relation.  This is shown in Figure~\ref{fig:lt} along with a
number of low-$z$ groups \citep{OsmondPonman2004} and clusters
\citep{Markevitch1998,Wuetal1999,Vikhlininetal2009a,Prattetal2009}, with
all values corrected with $E(z)$ scaling for self-similar evolution.  All
values have been scaled to $h = 0.7$ and are measured within or corrected
to \rfh\ with the exception of the \citet{Wuetal1999} clusters, which are
measured within 1 Mpc.  The FGs fall close to the locus of points traced by
low-$z$ systems, although they are consistent with rich groups and poor
\textit{clusters} instead of the poor group scales that one would infer
from their richness.  This is expected from previous observations of FGs
\citep[e.g.][]{Khosroshahietal2007} and indeed from the empirical
definition of a fossil group.  There is a hint from Figure~\ref{fig:lt}
that these FGs are either hotter or less luminous than normal systems,
although the errors are large.  The slightly larger aperture correction
favored by a $\beta = 0.67$, $\rc = 200$ kpc surface brightness model would
move the FG points closer in $L_X$ to the locus of points from normal
systems.

\begin{figure*}
\centering
\includegraphics[height=.9\linewidth,angle=270]{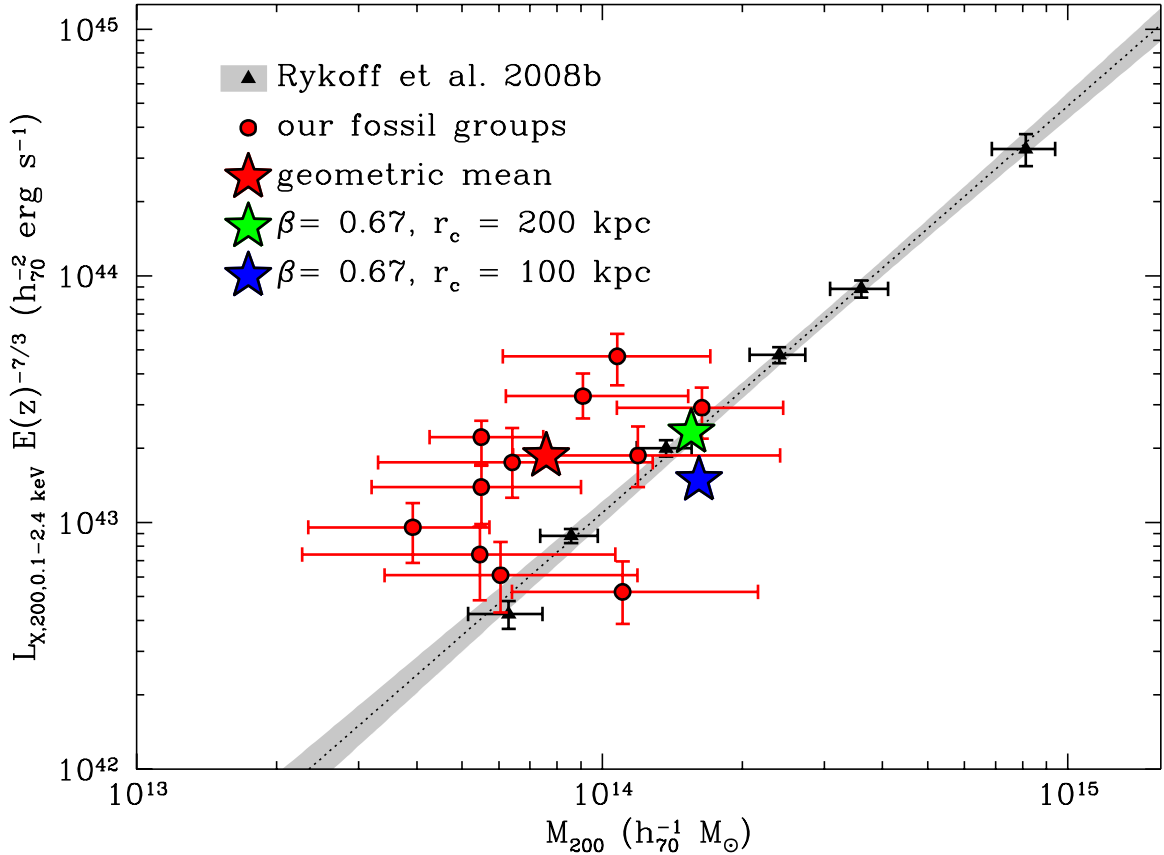}
\caption{$L_X$--$\mth$ relation from stacking of weak lensing measurements
of maxBCG clusters \citep{Rykoffetal2008b}, with our FGs overplotted.  The
red star shows the geometric mean of our systems; it is clearly offset to
lower mass and higher luminosity compared to the weak lensing results.
However, systematic error in our $\beta$-model surface brightness fitting
is large enough to account for this offset, as demonstrated by the other
star points, which show the centroid we would have obtained had we used
values of $\beta$ and $\rc$ typically found for normal clusters.}
\label{fig:ml}
\end{figure*}

With our modest X-ray data we can explore other scaling relations that
constrain different cluster characteristics.  The $L_X$--$M$ relation is a
useful probe of total baryon fraction, which is dominated by the hot ICM.
Using Eq.~\ref{eq:mass} for \rth, we calculate \mth\ for our FGs and
extrapolate the X-ray luminosity to \rth, using the method described in
Section~\ref{sect:spatial}.  The $L_X$--\mth\ relation is shown in
Figure~\ref{fig:ml}, along with the relation from \citet{Rykoffetal2008b}
for maxBCG clusters.  We correct for self-similar evolution at the redshift
$z = 0.25$ used in this work, which uses stacked weak lensing measurements
to estimate masses.  Our FGs fall above the \citet{Rykoffetal2008b} fit, at
higher luminosity for a given mass, which would indicate an enhanced baryon
fraction.  However, the systematic error due to $\beta$-model extrapolation
(described above) is much larger than this typical offset; as illustrated
in Figure~\ref{fig:ml}, this systematic effect is more likely to
underestimate the mass.  The FGs are therefore consistent with $L_X$--$M$
scaling relation of all maxBCG clusters, to the extent that we can
constrain the hydrostatic masses of our FGs.

\begin{figure*}
\centering
\includegraphics[height=.9\linewidth,angle=270]{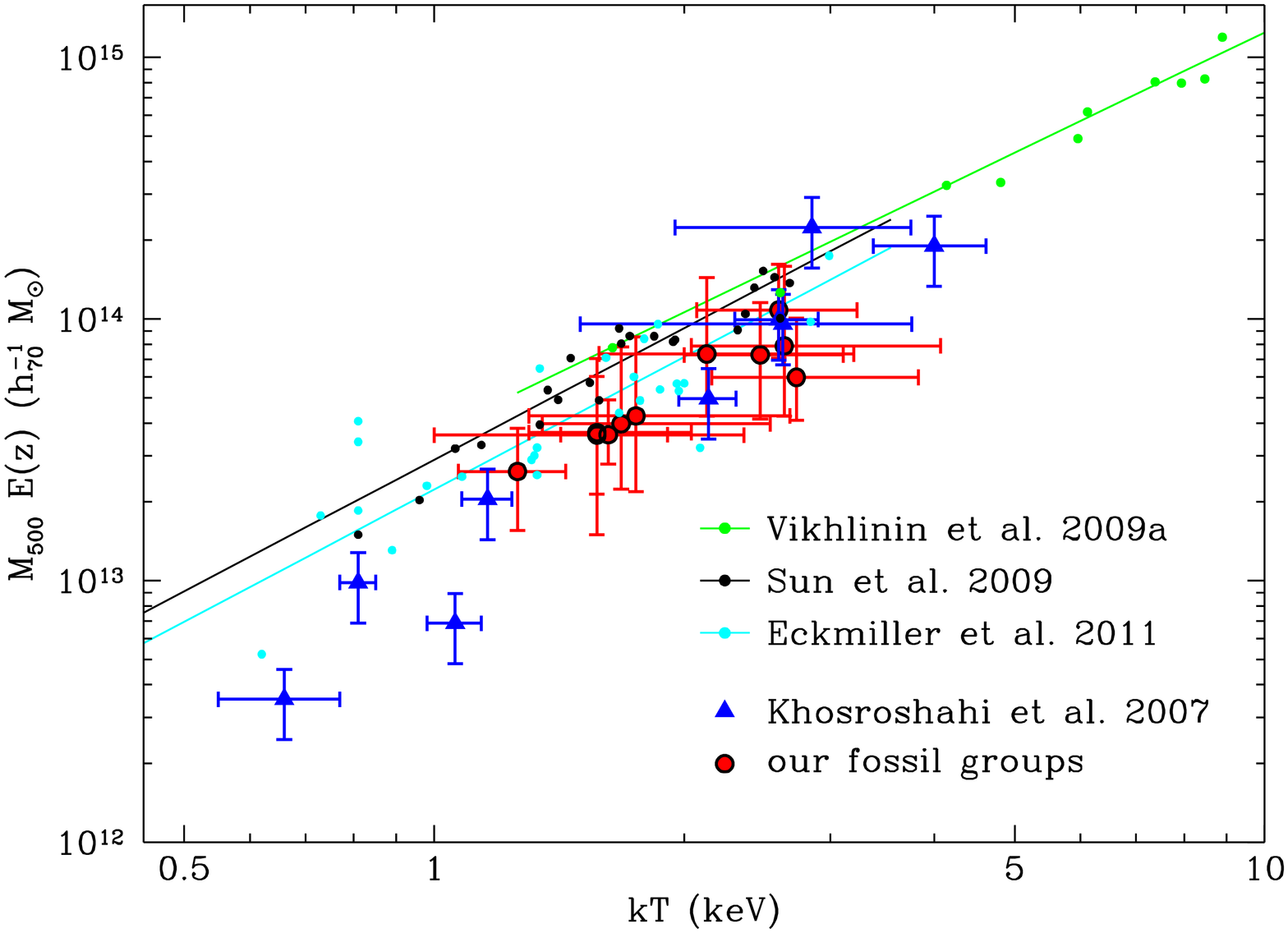}
\caption{$\mfh$--$T_X$ relation for our FGs, along with data and best-fit
trends for clusters \citep{Vikhlininetal2009a} and groups
\citep{Sunetal2009,Eckmilleretal2011} from the literature.
Also plotted are FGs and OLEGs from \citet{Khosroshahietal2007}, which form
a common locus with our own data and appear to be hotter for a given mass
than normal systems.  The systematic error of our mass estimates is larger
than the offset, as shown in Figure~\ref{fig:ml} and discussed in the
text.}
\label{fig:mt}
\end{figure*}

Likewise, we compare the $M$--$T_X$ relation to recent results from the
literature \citep{Vikhlininetal2009a,Sunetal2009,Eckmilleretal2011} that
have again been corrected for self-similar evolution (see
Figure~\ref{fig:mt}).  \citet{Khosroshahietal2007} suggest that their
sample of FGs and ``overluminous elliptical galaxies'' (OLEGs), also shown
in Figure~\ref{fig:mt}, are hotter for a given mass compared to normal
systems, an effect that increases toward lower mass.  They tentatively
attribute this to a lack of cool cores in the FGs driving the
emission-weighted temperature to higher values compared to normal systems.
We also find a systematically higher $T_X$ for a given mass in our sample;
however, the factor of $\sim 2$ systematic error in the mass is more than
adequate to explain this offset, and as shown in the $L_X$--$M$ relation,
it is more likely we have underestimated the masses than overestimated
them.  We note that \citet{Vikhlininetal2009a} and \citet{Sunetal2009}
exclude the core of the X-ray emission within 0.15\,\rfh\ in their
analyses, so cool cores (or lack thereof) cannot directly cause the
temperature differences seen in Figure~\ref{fig:mt}.  Our derived
$M$--$T_X$ for FGs is stil consistent with that of normal groups within the
relatively large statistical and systematic errors.

\subsection{Notes on Individual Systems}
\label{sect:indiv}

\noindent
\textit{J0133$-$1026}---The X-ray IGM of this system is the faintest in our
sample, and we are unable to constrain its temperature or mass.  In
contrast, the BCG contains a bright apparent point source with a flux
about one-third that of the thermal group emission.  This source is
well-fit by a power law with $\Gamma = 1.8^{+0.2}_{-0.4}$, as described
above, and we conclude that it is an AGN in the BCG.  
This conclusion is supported by two FIRST radio lobes projected 8\arcsec\
(17 kpc) and 11\arcsec\ (23 kpc) from the BCG center, with a total flux
density of 370 mJy.  While very bright, these sources were missed in our
filtering discussed in Section~\ref{sect:opt}, which only eliminated FIRST
sources within 3\arcsec.  The optical spectrum lacks emission
lines indicative of an AGN \citep{MauchSadler2007}.
The extended X-ray emission does not peak at the BCG but appears as a
clumpy ring out to a few hundred kpc.  The exclusion of the central point
source is unlikely to cause this structure, since the region we excluded is
quite small compared to the extent of the diffuse emission.  Deeper
observations will clarify the morphology and temperature of the IGM.

\noindent
\textit{J0815$+$3959}, \textit{J1039$+$3947},
\textit{J1411$+$5736}---These three systems share some morphological
traits.  The diffuse X-ray emission is not centrally concentrated in these
shallow observations, and each BCG is separated from the peak of the
diffuse X-ray emission by more than the uncertainty in the center of the
$\beta$ model fit (typically 6\arcsec).  J0815$+$3959 is devoid of an
obvious BCG X-ray source, while J1411$+$5736 has an extended BCG source
that produces 23\% of the flux of the IGM.  This latter source might simply
be the peak of the IGM, and its exclusion could explain the unusual
morphology seen in Figure~\ref{fig:smimages}.  J1039$+$3947 has a point-like BCG
X-ray source, as well as a 17 mJy FIRST source projected 3.2\arcsec\ (6 kpc) away,
just
missing our radio source filtering.  J1039$+$3947 fails the $\Delta_i > 2$
mag criterion using the \rth\ value derived from the temperature, 
$r_{200,kT} = 916$--1273 kpc (see Table~\ref{tab:r200}).  It meets this
criterion for the smaller value derived from the $\beta$-model fit; 
$r_{200,\beta} = 809$ kpc.

\noindent
\textit{J0906$+$0301}---As mentioned in Section~\ref{sect:success}, this
target was not detected in our 10 ksec \chandra\ observation, yet 25
galaxies around the BCG exhibit a similar redshift with a velocity
dispersion of $\sigma = 506\pm72$ \kps\ \citep{Proctoretal2011}.  From the
scaling relation presented by those authors, we infer $L_X \approx
3$--$30\eex{42}$ \ergss; this is mildly inconsistent with our 3$\sigma$
upper limit of $4.8\eex{42}$ \ergss.  Deeper observations of this target
will be illuminating, as it could be both optically and X-ray faint for its
inferred mass.

\noindent
\textit{J0856$+$0553},
\textit{J1007$+$3800},
\textit{J1410$+$4145}---Three of the four most X-ray luminous FGs, these
systems have well-constrained temperatures and $\beta$-model surface
brightness profiles.  The X-ray isophotes are regular for each FG, with the
optical BCG projected near the peak of the diffuse X-ray emission, although
J1007$+$3800 has somewhat compressed isophotes on the east side compared to
the west.
J0856$+$0553 contains no detected BCG X-ray source, while the other two
have extended BCG X-ray sources producing a small fraction of the total
X-ray flux.
J0856$+$0553 fails the $\Delta_i > 2$ mag criterion at the upper range of 
allowed \rth, but is otherwise consistent with the FG definition.

\noindent
\textit{J1017$+$0156}, \textit{J1133$+$5920}, \textit{J1153$+$6753}---These
systems appear relaxed and quite compact, with the detectable diffuse X-ray
emission all within a radius of 190 kpc.  All three
have X-ray centers within 6\arcsec\ (12 kpc) of the BCG location,
which is within the typical uncertainty for the center of the $\beta$ model
surface brightness fit.
J1153$+$6753 clearly has compact emission associated
with the BCG; we note that this target falls outside of the FIRST survey
field, but there are no other cataloged radio sources consistent 
with this position.
J1017$+$0156 has a more extended X-ray source associated with
the BCG. J1133$+$5920 has no obvious source in the center aside from
the extended IGM emission, although there are FIRST radio lobes projected
3\arcsec--4\arcsec\ (7--9 kpc) from the BCG, with a total flux density of
11 mJy.
With the updated \rth\ values shown in Table~\ref{tab:r200}, J1017$+$0156
strictly fails the $\Delta_i > 2$ mag criterion, with $\Delta_i = 1.8$.
Nevertheless, it is clearly an optically underluminous system 
\citep{Proctoretal2011}.

\noindent
\textit{J1045$+$0420}---This is the most X-ray luminous FG in the sample
and also the most distant.  The morphology of the diffuse emission is
strikingly irregular; the peak X-ray surface brightness is projected
1.2\arcmin\ (190 kpc) northwest of the BCG, and a fairly uniform bright
region extends from this point through the BCG to nearly \rfh\ in the
south.  A simple $\beta$ model centered on the BCG or X-ray centroid
is clearly not a representative surface brightness model, therefore the
estimated mass and extrapolation of the luminosity to \rfh\ are likely
incorrect.  There are two bright radio lobes projected $\sim$ 15\arcsec\
(40 kpc) east and $\sim$ 25\arcsec\ (67 kpc) west of the BCG, with a total
FIRST flux density of 160 mJy.  The irregular morphology and bright radio
lobes could be evidence of a recent merger or other activity.  Deeper X-ray
observations will help clarify this puzzling system.

\noindent
\textit{J1136$+$0713}---This group has a bright red-sequence member 0.5 $i$
mag fainter than the BCG and projected about 3.8\arcmin\ (440 kpc) to the
west.  Thus the $\Delta_i > 2$ mag criterion is not met at the upper range
of allowed \rth, although it is consistent at smaller \rth\ derived from
the $\beta$-model extrapolation.  Deeper X-ray observation will better
constrain \rth\ and determine whether this is a real FG.  The diffuse X-ray
emission is fairly regular and peaks at the location of the BCG.  No
separate BCG X-ray source is detected in this target.

\section{SUMMARY}
\label{sect:summary}

Fossil groups present a puzzle to current theories of structure formation.
Despite the low number of bright galaxies, their high velocity dispersions
\citep[e.g.,][]{Proctoretal2011} and high gas temperatures seem to indicate
cluster-like gravitational potential wells.  There have been very few FGs
with good quality X-ray data observed until recently, and their
idiosyncratic characteristics may contribute to enhance their apparent
contradictions.  We have embarked on a project to assemble a large sample
of optically identified FGs with a view toward dramatically increasing the
number of such systems with high quality X-ray data.

The principal observational results in this work are:

\noindent
(1) New \chandra\ X-ray detections were made for 12 new FGs,
from a sample of 15 optically selected groups from the maxBCG cluster
catalog with richness ranging from $9 \le N_{200} \le 25$.

\noindent
(2) The new X-ray data yielded temperatures for 11 of the FGs, ranging from
1.3 to 2.7 keV.  From these temperatures and an analysis of the surface
brightness profiles, we have estimated \rfh\ ranging from 440 to 710 kpc and
masses ranging from $M_{500} = 0.3$ to $1.0\eex{14}$ \msun.  These values of
masses and scaled radii are typical for groups and clusters in this
temperature range.

\noindent
(3) The $L_X$--$T_X$ relation for these new FGs does not deviate significantly
from the expectation for normal systems intermediate between
clusters and groups, although they tend to be more similar to galaxy
clusters. 

\noindent
(4) The $L_X$--$M$ and $M$--$T_X$ relations suggest that the FGs are on
average hotter and more luminous than normal systems, similar to the
results of \citet{Khosroshahietal2007}.  However, the systematic error
from luminosity correction and mass extrapolation are large enough to
explain these differences.

\noindent
(5) A small number (10--40\%) of the detected groups are morphologically
irregular, possibly due to past mergers, interaction of the IGM with a central
AGN, or superposition of multiple massive halos.  Two-thirds of the
X-ray-detected FGs exhibit X-ray emission associated with the central BCG,
although we are unable with the current data to distinguish between AGN and
extended thermal galaxy emission \citep[e.g., embedded galactic
coronae,][]{Sunetal2007}.

We conclude from these results that the selection criteria devised in
Section~\ref{sect:opt} were successful in finding real fossil groups, and
we have greatly increased the number of known fossil groups, a crucial step
for further statistical analysis.  The results obtained from 
further detailed studies of this sample will have implications for current
and future cluster population studies, and also for cosmology using galaxy
clusters.  The presence of a population of intermediate mass clusters with
small
numbers of galaxies may bias determinations of the mass function which
measure richness by galaxy counts.  This potentially biased mass function,
when used to set strong constraints on power spectrum normalization and
$\Omega_m$ \citep[e.g.,][and references therein]{Allenetal2011}, may in
turn bias these results.  Furthermore, such bias would also affect the
measurement of the mass function redshift evolution, which is used to
constrain the equation of state of dark energy \citep{Vikhlininetal2009b}.
In contrast to X-ray surveys, where good proxies are being refined very
quickly \citep[e.g.,][]{Kravtsovetal2006}, the largest current and
near-future cosmological surveys (e.g.,
DES\footnote{\url{http://www.darkenergysurvey.org}},
BOSS\footnote{\url{http://cosmology.lbl.gov/BOSS}}, 
J-PAS; \citeauthor{Benitezetal2009} \citeyear{Benitezetal2009}) estimate
the mass of clusters through optical mass proxies, using some type of
richness indicator
\citep[e.g.,][]{Rozoetal2009,Rozoetal2010,Rykoffetal2011}.  This dependence
on a possibly biased mass proxy drives the need to determine the mass range
of the effect, the relative abundance for this massive but optically poor
cluster population, and proper statistical correction methods for
cosmology.

With these data in hand, we have begun a systematic study of this sample.
The addition of optical radial velocities obtained for hundreds of galaxies
in these new fossil groups will allow us to study the scaling relations of
fossil groups presented in our companion paper \citep{Proctoretal2011}.  With
our planned deep X-ray follow-up with \xmm, we will better constrain the
IGM temperature, luminosity, metal abundance, and halo mass, and thoroughly
explore the morphology of the hot gas.  Finally, with additional proposed
\chandra\ snapshot observations, we will extend our sample to fainter
$L_{BCG}$, further testing the validity of our selection method and pushing
the scaling relations to lower masses.  

\acknowledgements
The authors would like to thank the referee for helpful comments that
improved the final manuscript.
Support for this work was provided by NASA through SAO Award Number
2834-MIT-SAO-4018 issued by the \chandra\ X-Ray Observatory Center, which
is operated by the Smithsonian Astrophysical Observatory for and on behalf
of NASA under contract NAS8-03060.  RD acknowledges additional financial
support from NASA Grant NNH10CD19C and partial support from \chandra\ Award
No.~GO9-0142A.  ESR thanks the TABASGO Foundation.  RLO acknowledges
financial support from the Brazilian agency FAPESP (Funda\c c\~ao de Amparo
\`a Pesquisa do Estado de S\~ao Paulo) through a Young Investigator Program
(numbers 2009/06295-7 and 2010/08341-3).

%%%%%%%%%%%%%%%%%%%%%%%%%%%%%%%%%%%%%%%%%%%%%%%%%%%%%%%%%%%%%%%%%%%%%%%%
%\bibliographystyle{apj}   % use apj.bst as the bibtex style file
%\bibliography{apj-jour,edm}

\begin{thebibliography}{72}
\expandafter\ifx\csname natexlab\endcsname\relax\def\natexlab#1{#1}\fi

\bibitem[{{Adelman-McCarthy} {et~al.}(2006){Adelman-McCarthy}, {Ag{\"u}eros},
  {Allam}, {Anderson}, {Anderson}, {Annis}, {Bahcall}, {Baldry}, {Barentine},
  {Berlind}, {Bernardi}, {Blanton}, {Boroski}, {Brewington}, {Brinchmann},
  {Brinkmann}, {Brunner}, {Budav{\'a}ri}, {Carey}, {Carr}, {Castander},
  {Connolly}, {Csabai}, {Czarapata}, {Dalcanton}, {Doi}, {Dong}, {Eisenstein},
  {Evans}, {Fan}, {Finkbeiner}, {Friedman}, {Frieman}, {Fukugita}, {Gillespie},
  {Glazebrook}, {Gray}, {Grebel}, {Gunn}, {Gurbani}, {de Haas}, {Hall},
  {Harris}, {Harvanek}, {Hawley}, {Hayes}, {Hendry}, {Hennessy}, {Hindsley},
  {Hirata}, {Hogan}, {Hogg}, {Holmgren}, {Holtzman}, {Ichikawa}, {Ivezi{\'c}},
  {Jester}, {Johnston}, {Jorgensen}, {Juri{\'c}}, {Kent}, {Kleinman}, {Knapp},
  {Kniazev}, {Kron}, {Krzesinski}, {Kuropatkin}, {Lamb}, {Lampeitl}, {Lee},
  {Leger}, {Lin}, {Long}, {Loveday}, {Lupton}, {Margon},
  {Mart{\'{\i}}nez-Delgado}, {Mandelbaum}, {Matsubara}, {McGehee}, {McKay},
  {Meiksin}, {Munn}, {Nakajima}, {Nash}, {Neilsen}, {Newberg}, {Newman},
  {Nichol}, {Nicinski}, {Nieto-Santisteban}, {Nitta}, {O'Mullane}, {Okamura},
  {Owen}, {Padmanabhan}, {Pauls}, {Peoples}, {Pier}, {Pope}, {Pourbaix},
  {Quinn}, {Richards}, {Richmond}, {Rockosi}, {Schlegel}, {Schneider},
  {Schroeder}, {Scranton}, {Seljak}, {Sheldon}, {Shimasaku}, {Smith}, {Smol{\v
  c}i{\'c}}, {Snedden}, {Stoughton}, {Strauss}, {SubbaRao}, {Szalay},
  {Szapudi}, {Szkody}, {Tegmark}, {Thakar}, {Tucker}, {Uomoto}, {Vanden Berk},
  {Vandenberg}, {Vogeley}, {Voges}, {Vogt}, {Walkowicz}, {Weinberg}, {West},
  {White}, {Xu}, {Yanny}, {Yocum}, {York}, {Zehavi}, {Zibetti}, \&
  {Zucker}}]{Adelman-McCarthyetal2006}
{Adelman-McCarthy}, J.~K., {et~al.} 2006, \apjs, 162, 38

\bibitem[{{Allen} {et~al.}(2006){Allen}, {Dunn}, {Fabian}, {Taylor}, \&
  {Reynolds}}]{Allenetal2006}
{Allen}, S.~W., {Dunn}, R.~J.~H., {Fabian}, A.~C., {Taylor}, G.~B., \&
  {Reynolds}, C.~S. 2006, \mnras, 372, 21

\bibitem[{{Allen} {et~al.}(2011){Allen}, {Evrard}, \& {Mantz}}]{Allenetal2011}
{Allen}, S.~W., {Evrard}, A.~E., \& {Mantz}, A.~B. 2011, \araa, 49, 409

\bibitem[{{Anders} \& {Grevesse}(1989)}]{AndersGrevesse1989}
{Anders}, E., \& {Grevesse}, N. 1989, \gca, 53, 197

\bibitem[{{Arnaud} \& {Evrard}(1999)}]{ArnaudEvrard1999}
{Arnaud}, M., \& {Evrard}, A.~E. 1999, \mnras, 305, 631

\bibitem[{{Balmaverde} {et~al.}(2008){Balmaverde}, {Baldi}, \&
  {Capetti}}]{Balmaverdeetal2008}
{Balmaverde}, B., {Baldi}, R.~D., \& {Capetti}, A. 2008, \aap, 486, 119

\bibitem[{{Becker} {et~al.}(2007){Becker}, {McKay}, {Koester}, {Wechsler},
  {Rozo}, {Evrard}, {Johnston}, {Sheldon}, {Annis}, {Lau}, {Nichol}, \&
  {Miller}}]{Beckeretal2007}
{Becker}, M.~R., {et~al.} 2007, \apj, 669, 905

\bibitem[{{Ben{\'{\i}}tez} {et~al.}(2009){Ben{\'{\i}}tez}, {Gazta{\~n}aga},
  {Miquel}, {Castander}, {Moles}, {Crocce}, {Fern{\'a}ndez-Soto}, {Fosalba},
  {Ballesteros}, {Campa}, {Cardiel-Sas}, {Castilla}, {Crist{\'o}bal-Hornillos},
  {Delfino}, {Fern{\'a}ndez}, {Fern{\'a}ndez-Sopuerta},
  {Garc{\'{\i}}a-Bellido}, {Lobo}, {Mart{\'{\i}}nez}, {Ortiz}, {Pacheco},
  {Paredes}, {Pons-Border{\'{\i}}a}, {S{\'a}nchez}, {S{\'a}nchez}, {Varela}, \&
  {de Vicente}}]{Benitezetal2009}
{Ben{\'{\i}}tez}, N., {et~al.} 2009, \apj, 691, 241

\bibitem[{{B{\"o}hringer} {et~al.}(2010){B{\"o}hringer}, {Pratt}, {Arnaud},
  {Borgani}, {Croston}, {Ponman}, {Ameglio}, {Temple}, \&
  {Dolag}}]{Bohringeretal2010}
{B{\"o}hringer}, H., {et~al.} 2010, \aap, 514, A32

\bibitem[{{Cash}(1979)}]{Cash1979}
{Cash}, W. 1979, \apj, 228, 939

\bibitem[{{Cui} {et~al.}(2011){Cui}, {Springel}, {Yang}, {De Lucia}, \&
  {Borgani}}]{Cuietal2011}
{Cui}, W., {Springel}, V., {Yang}, X., {De Lucia}, G., \& {Borgani}, S. 2011,
  \mnras, 416, 2997

\bibitem[{{Cypriano} {et~al.}(2006){Cypriano}, {Mendes de Oliveira}, \&
  {Sodr{\'e}}}]{Cyprianoetal2006}
{Cypriano}, E.~S., {Mendes de Oliveira}, C.~L., \& {Sodr{\'e}}, Jr., L. 2006,
  \aj, 132, 514

\bibitem[{{Dariush} {et~al.}(2010){Dariush}, {Raychaudhury}, {Ponman},
  {Khosroshahi}, {Benson}, {Bower}, \& {Pearce}}]{Dariushetal2010}
{Dariush}, A.~A., {Raychaudhury}, S., {Ponman}, T.~J., {Khosroshahi}, H.~G.,
  {Benson}, A.~J., {Bower}, R.~G., \& {Pearce}, F. 2010, \mnras, 405, 1873

\bibitem[{{D'Onghia} {et~al.}(2005){D'Onghia}, {Sommer-Larsen}, {Romeo},
  {Burkert}, {Pedersen}, {Portinari}, \& {Rasmussen}}]{DOnghiaetal2005}
{D'Onghia}, E., {Sommer-Larsen}, J., {Romeo}, A.~D., {Burkert}, A., {Pedersen},
  K., {Portinari}, L., \& {Rasmussen}, J. 2005, \apjl, 630, L109

\bibitem[{{Dupke} {et~al.}(2010){Dupke}, {Miller}, {de Oliveira}, {Sodre},
  {Rykoff}, {de Oliveira}, \& {Proctor}}]{Dupkeetal2010}
{Dupke}, R., {Miller}, E., {de Oliveira}, C.~M., {Sodre}, L., {Rykoff},
  E., {de Oliveira}, R.~L., \& {Proctor}, R. 2010, Highlights of Astronomy, 15,
  287

\bibitem[{{Eckmiller} {et~al.}(2011){Eckmiller}, {Hudson}, \&
  {Reiprich}}]{Eckmilleretal2011}
{Eckmiller}, H.~J., {Hudson}, D.~S., \& {Reiprich}, T.~H. 2011, \aap, in
press (arXiv:1109.6498)

\bibitem[{{Eigenthaler} \& {Zeilinger}(2009)}]{EigenthalerZeilinger2009}
{Eigenthaler}, P., \& {Zeilinger}, W.~W. 2009, Astronomische Nachrichten, 330,
  978

\bibitem[{{Finoguenov} \& {Ponman}(1999)}]{FinoguenovPonman1999}
{Finoguenov}, A., \& {Ponman}, T.~J. 1999, \mnras, 305, 325

\bibitem[{{Finoguenov} {et~al.}(2007){Finoguenov}, {Guzzo}, {Hasinger},
  {Scoville}, {Aussel}, {B{\"o}hringer}, {Brusa}, {Capak}, {Cappelluti},
  {Comastri}, {Giodini}, {Griffiths}, {Impey}, {Koekemoer}, {Kneib},
  {Leauthaud}, {Le F{\`e}vre}, {Lilly}, {Mainieri}, {Massey}, {McCracken},
  {Mobasher}, {Murayama}, {Peacock}, {Sakelliou}, {Schinnerer}, {Silverman},
  {Smol{\v c}i{\'c}}, {Taniguchi}, {Tasca}, {Taylor}, {Trump}, \&
  {Zamorani}}]{Finoguenovetal2007}
{Finoguenov}, A., {et~al.} 2007, \apjs, 172, 182

\bibitem[{{Hansen} {et~al.}(2005){Hansen}, {McKay}, {Wechsler}, {Annis},
  {Sheldon}, \& {Kimball}}]{Hansenetal2005}
{Hansen}, S.~M., {McKay}, T.~A., {Wechsler}, R.~H., {Annis}, J., {Sheldon},
  E.~S., \& {Kimball}, A. 2005, \apj, 633, 122

\bibitem[{{Helsdon} \& {Ponman}(2003)}]{HelsdonPonman2003}
{Helsdon}, S.~F., \& {Ponman}, T.~J. 2003, \mnras, 340, 485

\bibitem[{{Hickox} {et~al.}(2009){Hickox}, {Jones}, {Forman}, {Murray},
  {Kochanek}, {Eisenstein}, {Jannuzi}, {Dey}, {Brown}, {Stern}, {Eisenhardt},
  {Gorjian}, {Brodwin}, {Narayan}, {Cool}, {Kenter}, {Caldwell}, \&
  {Anderson}}]{Hickoxetal2009}
{Hickox}, R.~C., {et~al.} 2009, \apj, 696, 891

\bibitem[{{Jeltema} {et~al.}(2008){Jeltema}, {Hallman}, {Burns}, \&
  {Motl}}]{Jeltemaetal2008}
{Jeltema}, T.~E., {Hallman}, E.~J., {Burns}, J.~O., \& {Motl}, P.~M. 2008,
  \apj, 681, 167

\bibitem[{{Jeltema} {et~al.}(2006){Jeltema}, {Mulchaey}, {Lubin}, {Rosati}, \&
  {B{\"o}hringer}}]{Jeltemaetal2006}
{Jeltema}, T.~E., {Mulchaey}, J.~S., {Lubin}, L.~M., {Rosati}, P., \&
  {B{\"o}hringer}, H. 2006, \apj, 649, 649

\bibitem[{{Jeltema} {et~al.}(2009){Jeltema}, {Gerke}, {Laird}, {Willmer},
  {Coil}, {Cooper}, {Davis}, {Nandra}, \& {Newman}}]{Jeltemaetal2009}
{Jeltema}, T.~E., {et~al.} 2009, \mnras, 399, 715

\bibitem[{{Johnston} {et~al.}(2007){Johnston}, {Sheldon}, {Wechsler}, {Rozo},
  {Koester}, {Frieman}, {McKay}, {Evrard}, {Becker}, \&
  {Annis}}]{Johnstonetal2007}
{Johnston}, D.~E., {et~al.} 2007, \apj, submitted (arXiv:0709.1159)

\bibitem[{{Jones} \& {Forman}(1999)}]{JonesForman1999}
{Jones}, C., \& {Forman}, W. 1999, \apj, 511, 65

\bibitem[{{Jones} {et~al.}(2000){Jones}, {Ponman}, \& {Forbes}}]{Jonesetal2000}
{Jones}, L.~R., {Ponman}, T.~J., \& {Forbes}, D.~A. 2000, \mnras, 312, 139

\bibitem[{{Jones} {et~al.}(2003){Jones}, {Ponman}, {Horton}, {Babul},
  {Ebeling}, \& {Burke}}]{Jonesetal2003}
{Jones}, L.~R., {Ponman}, T.~J., {Horton}, A., {Babul}, A., {Ebeling}, H., \&
  {Burke}, D.~J. 2003, \mnras, 343, 627

\bibitem[{{Kalberla} {et~al.}(2005){Kalberla}, {Burton}, {Hartmann}, {Arnal},
  {Bajaja}, {Morras}, \& {P{\"o}ppel}}]{Kalberlaetal2005}
{Kalberla}, P.~M.~W., {Burton}, W.~B., {Hartmann}, D., {Arnal}, E.~M.,
  {Bajaja}, E., {Morras}, R., \& {P{\"o}ppel}, W.~G.~L. 2005, \aap, 440, 775

\bibitem[{{Kauffmann} {et~al.}(2003){Kauffmann}, {Heckman}, {Tremonti},
  {Brinchmann}, {Charlot}, {White}, {Ridgway}, {Brinkmann}, {Fukugita}, {Hall},
  {Ivezi{\'c}}, {Richards}, \& {Schneider}}]{khtbc03}
{Kauffmann}, G., {et~al.} 2003, \mnras, 346, 1055

\bibitem[{{Khosroshahi} {et~al.}(2004){Khosroshahi}, {Jones}, \&
  {Ponman}}]{Khosroshahietal2004}
{Khosroshahi}, H.~G., {Jones}, L.~R., \& {Ponman}, T.~J. 2004, \mnras, 349,
  1240

\bibitem[{{Khosroshahi} {et~al.}(2006{\natexlab{a}}){Khosroshahi}, {Maughan},
  {Ponman}, \& {Jones}}]{Khosroshahietal2006}
{Khosroshahi}, H.~G., {Maughan}, B.~J., {Ponman}, T.~J., \& {Jones}, L.~R.
  2006{\natexlab{a}}, \mnras, 369, 1211

\bibitem[{{Khosroshahi} {et~al.}(2006{\natexlab{b}}){Khosroshahi}, {Ponman}, \&
  {Jones}}]{KhosroshahiPonmanJones2006}
{Khosroshahi}, H.~G., {Ponman}, T.~J., \& {Jones}, L.~R. 2006{\natexlab{b}},
  \mnras, 372, L68

\bibitem[{{Khosroshahi} {et~al.}(2007){Khosroshahi}, {Ponman}, \&
  {Jones}}]{Khosroshahietal2007}
---. 2007, \mnras, 377, 595

\bibitem[{{Koester} {et~al.}(2007{\natexlab{a}}){Koester}, {McKay}, {Annis},
  {Wechsler}, {Evrard}, {Bleem}, {Becker}, {Johnston}, {Sheldon}, {Nichol},
  {Miller}, {Scranton}, {Bahcall}, {Barentine}, {Brewington}, {Brinkmann},
  {Harvanek}, {Kleinman}, {Krzesinski}, {Long}, {Nitta}, {Schneider},
  {Sneddin}, {Voges}, \& {York}}]{Koesteretal2007b}
{Koester}, B.~P., {et~al.} 2007{\natexlab{a}}, \apj, 660, 239

\bibitem[{{Koester} {et~al.}(2007{\natexlab{b}}){Koester}, {McKay}, {Annis},
  {Wechsler}, {Evrard}, {Rozo}, {Bleem}, {Sheldon}, \&
  {Johnston}}]{Koesteretal2007a}
---. 2007{\natexlab{b}}, \apj, 660, 221

\bibitem[{{Kravtsov} {et~al.}(2006){Kravtsov}, {Vikhlinin}, \&
  {Nagai}}]{Kravtsovetal2006}
{Kravtsov}, A.~V., {Vikhlinin}, A., \& {Nagai}, D. 2006, \apj, 650, 128

\bibitem[{{La Barbera} {et~al.}(2009){La Barbera}, {de Carvalho}, {de la Rosa},
  {Sorrentino}, {Gal}, \& {Kohl-Moreira}}]{laBarberaetal2009}
{La Barbera}, F., {de Carvalho}, R.~R., {de la Rosa}, I.~G., {Sorrentino}, G.,
  {Gal}, R.~R., \& {Kohl-Moreira}, J.~L. 2009, \aj, 137, 3942

\bibitem[{{Markevitch}(1998)}]{Markevitch1998}
{Markevitch}, M. 1998, \apj, 504, 27

\bibitem[{{Mauch} \& {Sadler}(2007)}]{MauchSadler2007}
{Mauch}, T., \& {Sadler}, E.~M. 2007, \mnras, 375, 931

\bibitem[{{Maughan} {et~al.}(2006){Maughan}, {Jones}, {Ebeling}, \&
  {Scharf}}]{Maughanetal2006}
{Maughan}, B.~J., {Jones}, L.~R., {Ebeling}, H., \& {Scharf}, C. 2006, \mnras,
  365, 509

\bibitem[{{Mendes de Oliveira} {et~al.}(2009){Mendes de Oliveira}, {Cypriano},
  {Dupke}, \& {Sodr{\'e}}}]{MendesdeOliveiraetal2009}
{Mendes de Oliveira}, C.~L., {Cypriano}, E.~S., {Dupke}, R.~A., \& {Sodr{\'e}},
  L. 2009, \aj, 138, 502

\bibitem[{{Mendes de Oliveira} {et~al.}(2006){Mendes de Oliveira}, {Cypriano},
  \& {Sodr{\'e}}}]{MendesdeOliveiraetal2006}
{Mendes de Oliveira}, C.~L., {Cypriano}, E.~S., \& {Sodr{\'e}}, Jr., L. 2006,
  \aj, 131, 158

\bibitem[{{Mittal} {et~al.}(2009){Mittal}, {Hudson}, {Reiprich}, \&
  {Clarke}}]{Mittaletal2009}
{Mittal}, R., {Hudson}, D.~S., {Reiprich}, T.~H., \& {Clarke}, T. 2009, \aap,
  501, 835

\bibitem[{{Mulchaey} \& {Zabludoff}(1999)}]{MulchaeyZabludoff1999}
{Mulchaey}, J.~S., \& {Zabludoff}, A.~I. 1999, \apj, 514, 133

\bibitem[{{Navarro} {et~al.}(1997){Navarro}, {Frenk}, \& {White}}]{NFW1997}
{Navarro}, J.~F., {Frenk}, C.~S., \& {White}, S.~D.~M. 1997, \apj, 490, 493

\bibitem[{{Osmond} \& {Ponman}(2004)}]{OsmondPonman2004}
{Osmond}, J.~P.~F., \& {Ponman}, T.~J. 2004, \mnras, 350, 1511

\bibitem[{{Ponman} {et~al.}(1994){Ponman}, {Allan}, {Jones}, {Merrifield},
  {McHardy}, {Lehto}, \& {Luppino}}]{Ponmanetal1994}
{Ponman}, T.~J., {Allan}, D.~J., {Jones}, L.~R., {Merrifield}, M., {McHardy},
  I.~M., {Lehto}, H.~J., \& {Luppino}, G.~A. 1994, \nat, 369, 462

\bibitem[{{Pratt} {et~al.}(2009){Pratt}, {Croston}, {Arnaud}, \&
  {B{\"o}hringer}}]{Prattetal2009}
{Pratt}, G.~W., {Croston}, J.~H., {Arnaud}, M., \& {B{\"o}hringer}, H. 2009,
  \aap, 498, 361

\bibitem[{{Proctor} {et~al.}(2011){Proctor}, {Mendes de Oliveira}, {Dupke},
  {Lopes de Oliveira}, {Cypriano}, {Miller}, \& {Rykoff}}]{Proctoretal2011}
{Proctor}, R.~N., {Mendes de Oliveira}, C.~L., {Dupke}, R.~A., {Lopes de
  Oliveira}, R., {Cypriano}, E.~S., {Miller}, E.~D., \& {Rykoff}, E.~S. 2011,
  \mnras, in press (arXiv:1108.1349)

\bibitem[{{Rasmussen} \& {Ponman}(2007)}]{RasmussenPonman2007}
{Rasmussen}, J., \& {Ponman}, T.~J. 2007, \mnras, 380, 1554

\bibitem[{{Rozo} {et~al.}(2007){Rozo}, {Wechsler}, {Koester}, {McKay},
  {Evrard}, {Johnston}, {Sheldon}, {Annis}, \& {Frieman}}]{Rozoetal2007a}
{Rozo}, E., {et~al.} 2007, arXiv:astro-ph/0703571

\bibitem[{{Rozo} {et~al.}(2009){Rozo}, {Rykoff}, {Koester}, {McKay}, {Hao},
  {Evrard}, {Wechsler}, {Hansen}, {Sheldon}, {Johnston}, {Becker}, {Annis},
  {Bleem}, \& {Scranton}}]{Rozoetal2009}
---. 2009, \apj, 703, 601

\bibitem[{{Rozo} {et~al.}(2010){Rozo}, {Wechsler}, {Rykoff}, {Annis}, {Becker},
  {Evrard}, {Frieman}, {Hansen}, {Hao}, {Johnston}, {Koester}, {McKay},
  {Sheldon}, \& {Weinberg}}]{Rozoetal2010}
---. 2010, \apj, 708, 645

\bibitem[{{Rykoff} {et~al.}(2008){Rykoff}, {McKay}, {Becker}, {Evrard},
  {Johnston}, {Koester}, {Rozo}, {Sheldon}, \& {Wechsler}}]{Rykoffetal2008}
{Rykoff}, E.~S., {et~al.} 2008, \apj, 675, 1106

\bibitem[{{Rykoff} {et~al.}(2008{\natexlab{b}}){Rykoff}, {Evrard}, {McKay},
  {Becker}, {Johnston}, {Koester}, {Nord}, {Rozo}, {Sheldon}, {Stanek}, \&
  {Wechsler}}]{Rykoffetal2008b}
---. 2008{\natexlab{b}}, \mnras, 387, L28

\bibitem[{{Rykoff} {et~al.}(2011){Rykoff}, {Koester}, {Rozo}, {Annis},
  {Evrard}, {Hansen}, {Hao}, {Johnston}, {McKay}, \&
  {Wechsler}}]{Rykoffetal2011}
---. 2011, \apj, submitted (arXiv:1104.2089)

\bibitem[{{Sheldon} {et~al.}(2009){Sheldon}, {Johnston}, {Scranton}, {Koester},
  {McKay}, {Oyaizu}, {Cunha}, {Lima}, {Lin}, {Frieman}, {Wechsler}, {Annis},
  {Mandelbaum}, {Bahcall}, \& {Fukugita}}]{Sheldonetal2009}
{Sheldon}, E.~S., {et~al.} 2009, \apj, 703, 2217

\bibitem[{{Smith} {et~al.}(2010){Smith}, {Khosroshahi}, {Dariush}, {Sanderson},
  {Ponman}, {Stott}, {Haines}, {Egami}, \& {Stark}}]{Smithetal2010}
{Smith}, G.~P., {et~al.} 2010, \mnras, 409, 169

\bibitem[{{Smith} {et~al.}(2001){Smith}, {Brickhouse}, {Liedahl}, \&
  {Raymond}}]{Smithetal2001}
{Smith}, R.~K., {Brickhouse}, N.~S., {Liedahl}, D.~A., \& {Raymond}, J.~C.
  2001, \apjl, 556, L91

\bibitem[{{Sun} {et~al.}(2004){Sun}, {Forman}, {Vikhlinin}, {Hornstrup},
  {Jones}, \& {Murray}}]{Sunetal2004}
{Sun}, M., {Forman}, W., {Vikhlinin}, A., {Hornstrup}, A., {Jones}, C., \&
  {Murray}, S.~S. 2004, \apj, 612, 805

\bibitem[{{Sun} {et~al.}(2007){Sun}, {Jones}, {Forman}, {Vikhlinin}, {Donahue},
  \& {Voit}}]{Sunetal2007}
{Sun}, M., {Jones}, C., {Forman}, W., {Vikhlinin}, A., {Donahue}, M., \&
  {Voit}, M. 2007, \apj, 657, 197

\bibitem[{{Sun} {et~al.}(2009){Sun}, {Voit}, {Donahue}, {Jones}, {Forman}, \&
  {Vikhlinin}}]{Sunetal2009}
{Sun}, M., {Voit}, G.~M., {Donahue}, M., {Jones}, C., {Forman}, W., \&
  {Vikhlinin}, A. 2009, \apj, 693, 1142

\bibitem[{{Vikhlinin}(2004)}]{Vikhlinin2004}
{Vikhlinin}, A. 2004, in \chandra\ Calibration Memo
  (\url{http://cxc.harvard.edu/contrib/alexey/contmap.pdf}: Cambridge: CXC)

\bibitem[{{Vikhlinin} {et~al.}(2009{\natexlab{a}}){Vikhlinin}, {Burenin},
  {Ebeling}, {Forman}, {Hornstrup}, {Jones}, {Kravtsov}, {Murray}, {Nagai},
  {Quintana}, \& {Voevodkin}}]{Vikhlininetal2009a}
{Vikhlinin}, A., {et~al.} 2009{\natexlab{a}}, \apj, 692, 1033

\bibitem[{{Vikhlinin} {et~al.}(2009{\natexlab{b}}){Vikhlinin}, {Kravtsov},
  {Burenin}, {Ebeling}, {Forman}, {Hornstrup}, {Jones}, {Murray}, {Nagai},
  {Quintana}, \& {Voevodkin}}]{Vikhlininetal2009b}
---. 2009{\natexlab{b}}, \apj, 692, 1060

\bibitem[{{von Benda-Beckmann} {et~al.}(2008){von Benda-Beckmann}, {D'Onghia},
  {Gottl{\"o}ber}, {Hoeft}, {Khalatyan}, {Klypin}, \&
  {M{\"u}ller}}]{vonBendaBeckmannetal2008}
{von Benda-Beckmann}, A.~M., {D'Onghia}, E., {Gottl{\"o}ber}, S., {Hoeft}, M.,
  {Khalatyan}, A., {Klypin}, A., \& {M{\"u}ller}, V. 2008, \mnras, 386, 2345

\bibitem[{{Wechsler} {et~al.}(2002){Wechsler}, {Bullock}, {Primack},
  {Kravtsov}, \& {Dekel}}]{Wechsleretal2002}
{Wechsler}, R.~H., {Bullock}, J.~S., {Primack}, J.~R., {Kravtsov}, A.~V., \&
  {Dekel}, A. 2002, \apj, 568, 52

\bibitem[{{White} {et~al.}(1997){White}, {Becker}, {Helfand}, \&
  {Gregg}}]{wbhg98}
{White}, R.~L., {Becker}, R.~H., {Helfand}, D.~J., \& {Gregg}, M.~D. 1997,
  \apj, 475, 479

\bibitem[{{Willis} {et~al.}(2005){Willis}, {Pacaud}, {Valtchanov}, {Pierre},
  {Ponman}, {Read}, {Andreon}, {Altieri}, {Quintana}, {Dos Santos},
  {Birkinshaw}, {Bremer}, {Duc}, {Galaz}, {Gosset}, {Jones}, \&
  {Surdej}}]{Willisetal2005}
{Willis}, J.~P., {et~al.} 2005, \mnras, 363, 675

\bibitem[{{Wu} {et~al.}(1999){Wu}, {Xue}, \& {Fang}}]{Wuetal1999}
{Wu}, X., {Xue}, Y., \& {Fang}, L. 1999, \apj, 524, 22

\bibitem[{{York} {et~al.}(2000){York}, {Adelman}, {Anderson}, {Anderson},
  {Annis}, {Bahcall}, {Bakken}, {Barkhouser}, {Bastian}, {Berman}, {Boroski},
  {Bracker}, {Briegel}, {Briggs}, {Brinkmann}, {Brunner}, {Burles}, {Carey},
  {Carr}, {Castander}, {Chen}, {Colestock}, {Connolly}, {Crocker}, {Csabai},
  {Czarapata}, {Davis}, {Doi}, {Dombeck}, {Eisenstein}, {Ellman}, {Elms},
  {Evans}, {Fan}, {Federwitz}, {Fiscelli}, {Friedman}, {Frieman}, {Fukugita},
  {Gillespie}, {Gunn}, {Gurbani}, {de Haas}, {Haldeman}, {Harris}, {Hayes},
  {Heckman}, {Hennessy}, {Hindsley}, {Holm}, {Holmgren}, {Huang}, {Hull},
  {Husby}, {Ichikawa}, {Ichikawa}, {Ivezi{\'c}}, {Kent}, {Kim}, {Kinney},
  {Klaene}, {Kleinman}, {Kleinman}, {Knapp}, {Korienek}, {Kron}, {Kunszt},
  {Lamb}, {Lee}, {Leger}, {Limmongkol}, {Lindenmeyer}, {Long}, {Loomis},
  {Loveday}, {Lucinio}, {Lupton}, {MacKinnon}, {Mannery}, {Mantsch}, {Margon},
  {McGehee}, {McKay}, {Meiksin}, {Merelli}, {Monet}, {Munn}, {Narayanan},
  {Nash}, {Neilsen}, {Neswold}, {Newberg}, {Nichol}, {Nicinski}, {Nonino},
  {Okada}, {Okamura}, {Ostriker}, {Owen}, {Pauls}, {Peoples}, {Peterson},
  {Petravick}, {Pier}, {Pope}, {Pordes}, {Prosapio}, {Rechenmacher}, {Quinn},
  {Richards}, {Richmond}, {Rivetta}, {Rockosi}, {Ruthmansdorfer}, {Sandford},
  {Schlegel}, {Schneider}, {Sekiguchi}, {Sergey}, {Shimasaku}, {Siegmund},
  {Smee}, {Smith}, {Snedden}, {Stone}, {Stoughton}, {Strauss}, {Stubbs},
  {SubbaRao}, {Szalay}, {Szapudi}, {Szokoly}, {Thakar}, {Tremonti}, {Tucker},
  {Uomoto}, {Vanden Berk}, {Vogeley}, {Waddell}, {Wang}, {Watanabe},
  {Weinberg}, {Yanny}, \& {Yasuda}}]{Yorketal2000}
{York}, D.~G., {et~al.} 2000, \aj, 120, 1579

\end{thebibliography}

\end{document}